\documentclass{aa}  

\usepackage{caption}
\usepackage{subcaption}
\usepackage{lipsum}
\usepackage{lscape}
\usepackage{amsmath}
\usepackage{xcolor}
\usepackage{booktabs}
\usepackage[symbol]{footmisc}
\usepackage{orcidlink} 
\usepackage{comment}
\usepackage{lastpage}

\usepackage{hyperref}
\hypersetup{
  colorlinks   = true, 
  urlcolor     = blue, 
  linkcolor    = blue, 
  citecolor   = blue 
}

\usepackage{graphicx}
\usepackage{txfonts}
\begin{document}

   \title{Constraining cosmological parameters using void statistics from the SDSS survey}

   \author{Elena Fernández-García\orcidlink{0009-0006-2125-9590}\thanks{e-mail: efdez$@$iaa.es}
          \inst{1} 
          \and
          Juan E. Betancort-Rijo
          \inst{2, 3}
          \and
          Francisco Prada\orcidlink{0000-0001-7145-8674}
          \inst{1}
          \and
          Tomoaki Ishiyama\orcidlink{0000-0002-5316-9171}
          \inst{4}
          \and
          Anatoly Klypin
          \inst{5}
          \and 
          Julia Ereza
          \inst{1}
          }

   \institute{Instituto de Astrofisica de Andalucia (CSIC), E18008 Granada, Spain,
         \and
             Instituto de Astrofisica de Canarias, C/ Via Lactea s/n, Tenerife E38200, Spain
            \and 
            Facultad de Fisica, Universidad de La Laguna, Astrofisico Francisco Sanchez, s/n, La Laguna, Tenerife E38200, Spain
            \and 
            Digital Transformation Enhancement Council, Chiba University, 1-33, Yayoi-cho, Inage-ku, Chiba, 263-8522, Japan
            \and
            Department of Astronomy, University of Virginia, Charlottesville, VA 22904, USA
             }

   \date{Received ..., ...; accepted ..., ...}

 
  \abstract
   {}
   {We aim to constrain the amplitude of the linear spectrum of density fluctuations ($\sigma_{8}$), the matter density parameter ($\Omega_{\rm m}$), the Hubble constant (H$_{0}$), $\Gamma=\Omega_{\rm c}h$, and S$_{8}$ from  Sloan Digital Sky Survey Data Release 7 (SDSS DR7) by studying the abundance of large voids in the large-scale structure of galaxies.}
   {Voids are identified as maximal non-overlapping spheres within SDSS DR7 galaxies with redshifts of $0.02<z<0.132$ and absolute magnitudes of $M_{r}<-20.5$. We used the theoretical framework developed in previous works and recalibrated the data using halo simulations  to constrain $\sigma_{8}$, $\Omega_{\rm m}$, and H$_{0}$ from the sample of SDSS galaxies mentioned above using a Bayesian analysis and Markov chain Monte Carlo (MCMC) technique. This method has also been validated using simulated halo boxes and galaxy lightcones.
   }
   {We have proven that the theoretical framework recovers $\sigma_{8}$, $\Omega_{\rm m}$, and H$_{0}$ values from the halo simulation boxes for different values of $\sigma_{8}$ within 1$\sigma$ ($2\sigma$) in real (redshift) space. The theoretical framework void statistics from mock lightcones shows significant potential: we have studied the marginalised posteriors in each plane and checked that we were able to recover Planck values for the all the parameters. The results we obtained from the SDSS sample are: $\sigma_{8}=1.520^{+0.416}_{-0.441}$, $\Omega_{\rm m}=0.459^{+0.184}_{-0.184}$, H$_{0}=71.63^{+12.60}_{-11.77}$, $\Gamma=0.270^{+0.094}_{-0.100}$, and S$_{8}=1.87^{+0.59}_{-0.76}$. Combining these constraints with the Kilo Degree Survey (KiDS-1000) and the Dark Energy Survey (DESY3) yields $\sigma_{8}=0.858^{+0.040}_{-0.040}$, $\Omega_{\rm m}=0.257^{+0.023}_{-0.020}$, H$_{0}=74.17^{+4.66}_{-4.66}$, and S$_{8}=0.794^{+0.016}_{-0.016}$. The combined uncertainties of $\sigma_{8}$ and $\Omega_{\rm m}$ have been reduced by a factor of 2-3, compared to KiDS-100+DESY3 alone, due to the nearly orthogonal marginalised posteriors of SDSS voids and weak lensing in the $\sigma_{8}-\Omega_{\rm m}$ plane.

   }
   {}


   \maketitle
%

\section{Introduction}
For over 40 years, it has been observed that the brightest galaxies are generally found in dense regions and that most of cosmic space is devoid of these types of galaxies. This distribution is amply understood as a natural evolution of density fluctuations in matter that have been progressively amplified by gravitational instability \citep{giovanelli2009void}. In this framework, it has been demonstrated that initially under-dense regions grow in size, while highly dense areas end up collapsing under their own gravity \citep{10.1111/j.1365-2966.2004.07661.x}.

Over the last century, there has been a significant focus on studying the over-dense regions of the Universe \cite[e.g.][]{1969MNRAS.143..129K, 1977ARA&A..15..505B, 1987MNRAS.227....1K, 1994MNRAS.269..301E, 2001ApJ...560L.111H, 2005MNRAS.357..608Y, 2006MNRAS.368...21L, 2007MNRAS.377L...5G, 2019MNRAS.485.1196Z, sdssmocks}, while under-dense regions only started attracting interest  more recently \cite[e.g.][]{1982ApJ...262L..23H, 1982Natur.300..407Z, 1992ApJ...388..234B, Little_1994, Goldwirth_1995, 10.1111/j.1365-2966.2004.07661.x, 2012MNRAS.421.3481L, 2014MNRAS.443.3238P, 2016MNRAS.455.3075P, Achitouv_2019, Chan2019, rodríguezmedrano2023evolutionary, curtis2024properties}. Studying these regions (voids) is very useful as they possess unique characteristics that make them important probes for cosmological studies and the physics of galaxy formation. They are useful, for example, in: 

\begin{itemize}
    \item constraining the equation of state for dark energy \cite[e.g.][]{leepark, 2010PhRvD..82b3002B, nbody, Contarini_2022},
    \item studying modified gravity \cite[e.g.][]{martino2009density, Clampitt_2013, Voivodic_2017, Falck_2017, Perico_2019, 2021MNRAS.504.5021C, Mauland_2023}
    \item constraining cosmological models \cite[e.g.][]{ryden, benson, croton, lavaux}, 
    \item constraining cosmological parameters based on their statistics \cite[e.g][]{art5, 10.1093/mnras/stw1340, Hamaus_2020, 10.1093/mnras/stac828, Contarini_2022, Contarini_2023},
    \item testing the primordial non-Gaussianities \cite[e.g.][]{2009ApJ...701L..25S, 2010ApJ...724..285C, Chan_2019},
    \item studying massive neutrinos and dark energy \citep{Verza_2019, Contarini_2022, Verza_2023, 2023arXiv230707555T}.
\end{itemize}

In fact,  any cosmological parameter could be constrained, in principle, from void statistics. Specifically, the abundance of voids with radii larger than $r$, $N_{v}(r)$ depends on the normalisation of the linear spectrum of density fluctuations ($\sigma_{8}$). The shape of this function for large values of $r$ depends on $\Gamma=\Omega_{\rm c}h$, where $h$=H$_{0}/100$ $kms^{-1}Mpc^{-1}$  (see \cite{art5} for details).

Constraining $\sigma_{8}$ and $\Gamma=\Omega_{\rm c}h$ is especially interesting as there  some statistically significant tension does exist for the cosmological parameter S$_{8}=\sigma_{8}\sqrt{\Omega_{\rm m}/0.3}$ between the Planck experiment \citep{planck2018}, which measures the cosmic microwave background (CMB) anisotropies (S$_{8}=0.834\pm0.0161$), and other low-redshift cosmological probes, such as weak gravitational lensing, where a value of S$_{8}=0.776^{+0.032}_{-0.030}$ was obtained by  the Kilo Degree Survey (KiDS-1000)  \citep{Li_2023}. This tension is known as the S$_{8}$ tension. (see \cite{Di_Valentino_2021} for more details).

Early studies of cosmic voids have been limited by the tiny areas inspected by the available surveys. Nevertheless, with the emergence of more expansive redshift surveys, such as  Two-Degree Field Galaxy Redshift Survey (2dFGRS) \citep{10.1046/j.1365-8711.2001.04902.x, colless20032df},  Sloan Digital Sky Survey (SDSS) \citep{2000AJ....120.1579Y} and the SDSS Baryon Oscillation Spectroscopic Survey (BOSS) \citep{2013AJ....145...10D}, alongside improved resolution in cosmological simulations and enhanced analytical methodologies, we now have the capability to derive precise statistical insights concerning voids \cite[e.g.][]{Tikhonov_2006, Ceccarelli_2006, vonbendabeckmann2007void, Tikhonov_2007, Patiri_2012, Hamaus_2020, Douglass_2023, Contarini_2023}.

However, despite the longstanding presence of the concept of cosmic voids, there is no universally accepted definition for what constitutes a void. The term "void" can encompass a range of disparate entities, depending on the data employed and the objectives of the analysis. For example, voids can be defined as under-dense regions based on the smoothed dark matter (or halo/galaxy) density field \citep{10.1111/j.1365-2966.2005.09064.x}, as gravitationally expanding regions based on the dynamics of the dark matter density field \citep{vol3, 2015MNRAS.448..642E} or as empty spatial regions among discrete tracers \citep{hagai, 1998ApJ...497..534A, hoyle2}.

The choice of a simple definition of voids, in particular, their definition as empty spheres, is convenient for statistical studies of galaxy voids. In this paper, we define voids as maximal non-overlapping spheres empty of objects with mass (or luminosity) above a given value. With this definition, it is clear that voids are not empty, as there can be low luminosity galaxies (or low mass halos) inside them.

The aim of this paper is to infer the values of $\sigma_{8}$, H$_{0}$, and $\Omega_{\rm m}$ from SDSS redshift survey using the theoretical framework of voids statistics developed in \cite{1990MNRAS.246..608B}. This has been already done in previous articles in different ways. For example, in \cite{Sahl_n_2016} galaxy cluster and void abundances are combined using extreme-value statistics on a large cluster and a void. In this way, they have been able to obtain a value of $\sigma_{8}=0.95\pm0.21$ for a flat $\Lambda CDM$ universe. In \cite{2016PhRvL.117i1302H}, the authors constrained $\Omega_{\rm m}=0.281\pm0.031$ by studying void dynamics in SDSS. In \cite{2023MNRAS.523.6360W} they also used SDSS survey to constrain $\Omega_{\rm m}=0.391^{+0.028}_{-0.021}$ by measuring the void-galaxy and galaxy-galaxy clustering. In \cite{ Contarini_2023} they modelled the void size function \citep{1974ApJ...187..425P, 10.1111/j.1365-2966.2004.07661.x} by means of an extension of the popular volume-conserving model \citep{10.1093/mnras/stt1169}, which is based on two additional nuisance parameters. Applying a Bayesian analysis, they obtained a value of $\sigma_{8}=0.79^{+0.09}_{-0.08}$ for the BOSS DR12 survey; in a posterior work \citep{Contarini_2024}, they constrained S$_{8}=0.813^{+0.093}_{-0.068}$ and H$_{0}=67.3^{+10}_{-9.1}$ for the same redshift survey. However, in these works, the uncertainties obtained for $\sigma_{8}$ and H$_{0}$ are very large and the constrained values are compatible with high- and low-redshift cosmological probes. In addition, none of these studies have sought to combine their results with the CMB or with weak lensing results. Therefore, in this work, we propose using a original theoretical framework,  presented for the first time in \cite{art5}, which we have calibrated and tested on different halo simulation boxes in this work. We inferred $\sigma_{8}$, $\Omega_{\rm m}$, and H$_{0}$ from SDSS galaxies with redshifts of $0.02<z<0.132$ and absolute magnitudes in the r-band of $M_{r}<-20.5.$  We combined these results with KiDS-1000 and the Dark Energy Survey (DESY3) weak lensing results.

This paper is structured as follows. In Section \ref{sec4}, we describe the redshift survey as well as the simulations used in this work. In Section \ref{sec5}, we give a detailed explanation of how our 'Void Finder' works and give the relevant statistics of voids obtained for the observational catalogue, comparing it with the result of Uchuu-SDSS lightcones. In Section \ref{sec2}, we introduce the most important concepts and equations used in the theoretical framework used to calculate the abundance of voids larger than $r$ and the void probability function (VPF). We demonstrate that the theoretical framework  successfully predicts these two void statistics for halo simulation boxes with different $\sigma_{8}$ values in real and redshift space, and we show the results for SDSS and the Uchuu-SDSS lightcones. In Section \ref{likelihoodtest}, we explain the statistical test we used to infer $\sigma_{8}$, $\Omega_{\rm m}$, and $h$. In Section \ref{SDSS_como_constraints}, we show the constrained values for the sample of Uchuu-SDSS lightcones and SDSS redshift survey we  used in this work and we combined our results with KiDS-1000 results \citep{2023OJAp....6E..36D}. In Section \ref{comparison_sec}, we compare our constrained values of $\sigma_{8}$, $\Omega_{\rm m}$, and H$_{0}$ with those obtained in other works where voids are also used. Finally, in Section \ref{discussion}, we summarise the most important results obtained in this work.

\section{Data and mocks}\label{sec4}
The aim of this work is to infer $\sigma_{8}$, $\Omega_{\rm m}$, and H$_{0}$ values from Sloan Digital Sky Survey. However, to make sure that the theoretical equations  correctly reproduce the void functions of this redshift survey, we also used four halo simulation boxes with different $\sigma_{8}$ values (see Table \ref{parametros_3}), one simulation galaxy box and 32 lightcones. In this section, we introduce the redshift survey and mocks.

\subsection{SDSS}
\label{sec:maths} 
We  used the seventh release \citep{sdss} of Sloan Digital Sky Survey (SDSS DR7) \citep{2000AJ....120.1579Y}, which includes 11663 $deg^{2}$ of CCD imaging data in five photometric bands for 357 million distant objects. The catalogue also completed spectroscopy over {9380} $deg^{2}$. In total, there are 1.6 million spectra, including {930000} galaxies, {120000} quasars, and {460000} stars.

However, in this work we use a subcatalogue of SDSS: only galaxies from the northern regions with completeness greater than 90$\%$ are selected. Therefore, the effective area of this subcatalogue is {6511} $deg^{2}$ and contains around {497000} galaxies with redshifts between $z \in (0,0.5)$. In this sample, $\sim$ 6$\%$ of targeted galaxies lack a spectroscopically measured redshift due to fibre collisions. Thus, a nearest neighbour correction was applied to these galaxies, assigning to them the redshift of the galaxy they have collided with \citep{sdssmocks}.

Additionally, we imposed further cuts on absolute magnitude and redshift. We constructed a volume-limited sample by keeping only galaxies brighter than the Milky Way-like galaxies ($M_{r}<-20.5$, where $M_{r}$ is the absolute magnitude in $r$-band) and with  $z \leq 0.132$. We additionally imposed $z\geq0.02,$ so that we did not take into account any nearby galaxies highly affected by peculiar velocities. The physical volume of this sample is $V=41.67\times10^{6} h^{-3}$Mpc$^{3}$. There is a total of 112496 galaxies that fulfill these restrictions, so the number density of the galaxies is $n_{g}=2.838\times10^{-3} h^{3}$Mpc$^{-3}$.

\subsection{Mocks}\label{mocks}
\begin{table}[tb!]
\caption{Cosmological information about the halo boxes used in this work.}
\centering
 \begin{tabular}{lccc}  \toprule
     &  Uchuu & P18(Low) & P18VeryLow \\
  \midrule
$\Omega_{\rm m}$  & 0.3089 & 0.3111 & 0.3111 \\
      $\Omega_{\Lambda}$ & 0.6911 & 0.6889 & 0.6889  \\ 
      $\Omega_{\rm b}$  &  0.0486 & 0.048975 & 0.048975 \\
      $h$  &  0.6774 & 0.6766 & 0.6766   \\
      $\sigma_{8}$ & 0.8159 & 0.8102 (0.75) &  0.65  \\
      $L_{\rm box}$ [$h^{-1}$Mpc] & 2000 & 1000 & 1000 \\
      $N_{\rm part}$ & $12800^{3}$ & $6400^{3}$ & $3200^{3}$ \\
      $M_{\rm part}$ $[h^{-1}\rm M_{\odot}]$ & $3.27\times10^{8}$ & $3.29\times10^{8}$ & $2.63\times 10^{9}$  \\
      $\varepsilon$ $[h^{-1}$kpc$]$ & 4.27 & 1.0 & 4.0 \\
      Haloes  & Yes & Yes & Yes  \\
      Galaxies & Yes & No & No  \\
      Lightcones & Yes (32) & No & No \\
  \bottomrule
 \end{tabular}
  \tablefoot{Cosmological parameters (first five rows), the size of the simulation box ($L_{\text{box}}$), the number of dark matter particles used in the simulation ($\rm N_{\text{part}}$), their mass ($M_{\text{part}}$), and the gravitational softening ($\varepsilon$) for Uchuu, P18, P18Low and P18VeryLow  used to generate the box catalogues in this work. The last five rows provide information about whether the box catalogue, galaxy box or lightcones for SDSS are available for each simulation. 
  }
\label{parametros_3}
\end{table}
We used the Uchuu simulation, which was produced using the \textit{TreePM} code \textit{GreeM}~\footnote[1]{\url{http://hpc.imit.chiba-u.jp/~ishiymtm/greem/}} \citep{art3, art4} on the supercomputer ATERUI II at Center for Computational Astrophysics, CfCA, of National Astronomical Observatory of Japan. The number of dark matter particles was {12800}$^{3}$ with a mass resolution of $3.27 \times 10^{8}h^{-1}M_{\odot}$ in a box with a side length of {2000} $h^{-1}$Mpc. A total of 50 halo catalogues (snapshots) were created, covering the redshift range from 0 to 14 \citep{ishiyama}. The halos were subsequently identified using the \textit{RockStar} halo/subhalo finder~\footnote[2]{\url{https://bitbucket.org/gfcstanford/rockstar/}} \citep{Behroozi2013} and merger trees were generated using the consistent trees code\footnote[3]{\url{https://bitbucket.org/pbehroozi/consistent-trees/}}
 \citep{Behroozi2013b}. These simulations adopted the cosmological parameters from Planck 2015 (see Table \ref{parametros_3}) \citep{planck2015}. All this data is publicly available and accessible in the \textit{Skies $\&$ Universes} database\footnote[4]{\url{http://www.skiesanduniverses.org/Simulations/Uchuu/}}, including galaxy catalogues constructed using various methods~\citep{Aung2023,Oogi2023,Gkogkou2023,Prada2023,Ereza2023,sdssmocks}. 

Apart from Uchuu, we use three more simulations which are being used for the first time in this work. One of these simulations have Planck 2018 cosmological parameters \citep{planck2018} P18 and the other two have Planck 2018 parameters, but with $\sigma_{8} = 0.75$ (P18LowS8) $\sigma_{8} = 0.65$ (P18VeryLowS8). Additionally, P18 and P18LowS8 have exactly the same simulation properties, except for the value of $\sigma_{8}$. P18VeryLowS8 has different value of $\sigma_{8}$ and different mass resolution. All these details can be seen in Table \ref{parametros_3}. These three simulations were produced using the \textit{TreePM} code \textit{GreeM} on the supercomputer Fugaku at the RIKEN Center for Computational Science. As well as the Uchuu simulation, we generated initial conditions for these three simulations with 2nd order Lagrangian perturbation theory \citep{Crocce2006} by \textit{2LPTIC} code~\footnote[5]{\url{http://cosmo.nyu.edu/roman/2LPT/}}. The initial conditions are identical across the three simulations, enabling us to compare them without cosmic variance. The total of 70 halo catalogues were created, where the redshift list is the same with the Uchuu at $z<6$. N-body data, including the halo catalogues and the merger trees are also available in the \textit{Skies $\&$ Universes} database.

\begin{figure}[tph!]
    \centering
    \includegraphics[width=0.5\textwidth]{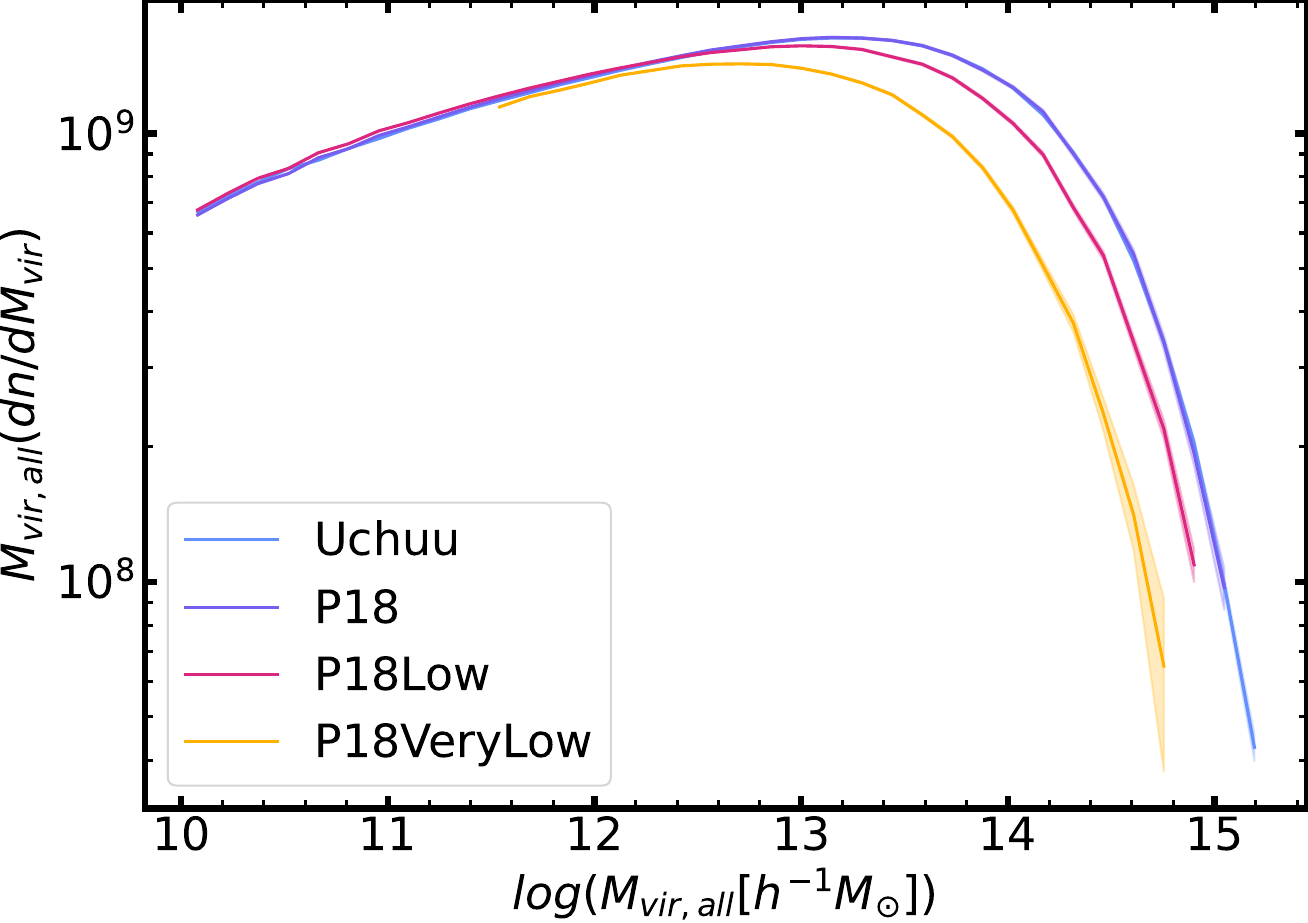}
    \caption{Halo mass function multiplied by the mean halo mass within virial radius (including unbound particles) of each bin of the four box catalogues used in this work. Shaded regions represent poissonian errors.
    }
    \label{hmf}
\end{figure}

The halo mass function (HMF) of each simulation can be seen in Figure \ref{hmf}, where $M_{vir, all}$ is the halo mass within virial radius including unbound particles. In this figure, it can be seen that VeryLow has lower mass resolution, as the HMF decreases considerably for small masses. It is also shown that (especially for large masses) the HMF power decreases proportionally with $\sigma_{8}$. Additionally, it can be seen that there is no significant difference between P18, Low and VeryLow halo mass functions for low virial masses ($M_{vir,all}<10^{12} h^{-1}M_{\odot}$), but there is a significant difference for large virial masses ($M_{vir,all}>10^{12} h^{-1}M_{\odot}$).

We also used a galaxy box with SDSS properties and 32 Uchuu-SDSS lightcones constructed with these galaxy boxes at different snapshots. These products are presented and extensively detailed in \cite{sdssmocks}. The galaxy box has been generated from the Uchuu simulation using the subhalo abundance matching (SHAM) method. Lightcones with SDSS properties have been constructed from the Uchuu-SDSS galaxy boxes, using a total of six snapshots between z = 0 and z = 0.5, which are separated in redshift by approximately 0.1 (z$_{\rm snap}$ = 0, 0.093, 0.19, 0.3, 0.43, and 0.49). We refer to \cite{sdssmocks} for more information about the methodology followed to construct the lightcones.

For the purpose of studying void statistics in simulation boxes, we select those snapshots corresponding to a redshift of $z\sim0.092$ (which is the snapshot with the closest redshift to SDSS median redshift). Moreover, we did not use all the objects (halos or galaxies) from this boxes, but we select a number density of $\bar{n}=3\times 10^{-3} h^{3}$Mpc$^{-3}$ for each box. This can be reached removing halos less massive than $M_{vir, all}$ = 1.626$\times 10^{12}M_{\odot}/h$ (Uchuu), 1.642$\times 10^{12}M_{\odot}/h$ (P18), 1.595$\times 10^{12}M_{\odot}/h$ (P18LowS8), and 1.457$\times 10^{12}M_{\odot}/h$ (P18VeryLow), along with galaxies that are less bright than $M_{r}$<-20.5 for Uchuu galaxy box.

The number density of the boxes, $\bar{n} = 3 \times 10^{-3} h^{3}$ Mpc$^{-3}$, does not exactly match the number density of SDSS or Uchuu-SDSS due to the incompleteness of the latter. However, by applying the same cut in $M_{r}$, we could ensure that the objects in the boxes are consistent with those in Uchuu-SDSS, except for those "lost" due to incompleteness.
Additionally, halo boxes were used to calibrate the theoretical framework and test its ability to successfully predict void functions for different values of $\sigma_{8}$.

\section{Statistics of voids in the SDSS}\label{sec5}

The next important step we needed to take once  the sample of halos or galaxies we are going to use is well defined is to then identify voids, defined as maximum non-overlapping spheres. There are a number of previous works that have used the same definition of voids and have developed their own void finder, such as \citep{10.1093/mnras/stt1169, 2017A&A...607A..24R, Contarini_2024, art1}.  In this work, we  use the same procedure as in \cite{art1}, namely: 

\begin{enumerate}
    \item We performed a Delaunay triangulation (with periodic conditions for simulation boxes), 
    \item Delaunay triangulation provides us with a list of spheres that fulfill the Delaunay condition:  the Delaunay triangulation of a set of points, ${p_{i}}$, ensures that no point, $p_{i}$, lies within the circumcircle of any triangle in the triangulation. However, these spheres can overlap, so we write an additional code that find voids among these spheres. 
\end{enumerate}

In order to calculate the Delaunay triangulation, we use the public code called Delaunay trIangulation Void findEr (\texttt{DIVE}\footnote[7]{\url{https://github.com/cheng-zhao/DIVE}}), which is introduced in \cite{art1}. This code uses  The Computational Geometry Algorithms Library\footnote[6]{\url{https://doc.cgal.org/4.6.3/Manual/packages.html\#PkgTriangulation3Summary}}, and, as mentioned above, the output of this code is a file that contains the positions in space of the centres of the spheres that satisfy the Delaunay condition, as well as their radii. However, these spheres are not voids, but candidates to be voids. 

To find voids (i.e. maximal non-overlapping spheres) among these spheres, an additional code needs to be developed. This code must check if two spheres overlap and, in case they do, keep the largest one as a void (see Appendix \ref{voidfinderalgorithm} for a more detailed explanation of the algorithm).

\begin{figure}[tbh!]
    \centering
    \includegraphics[width=0.5\textwidth]{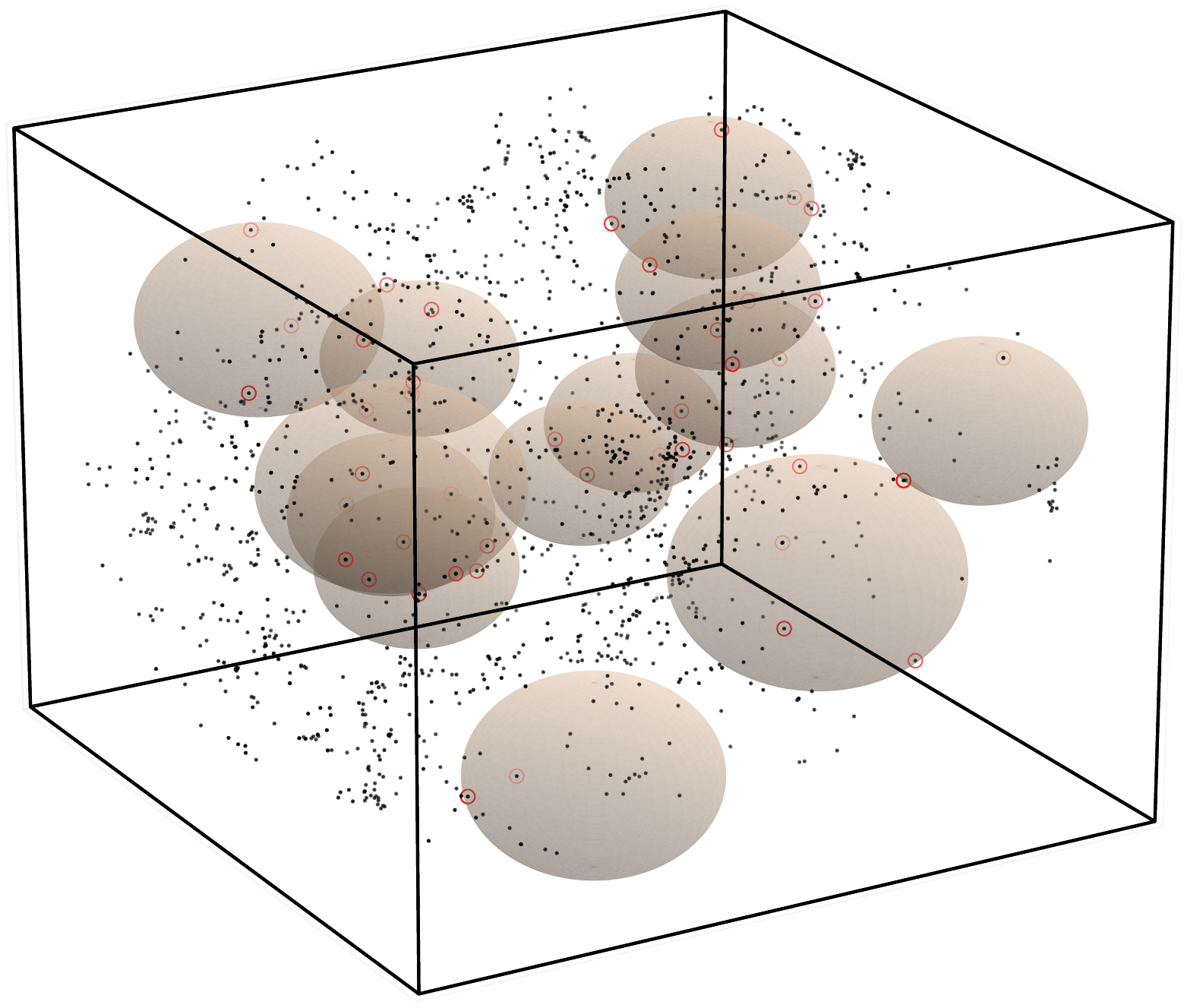}
    \caption{Voids with $r>9h^{-1}$Mpc (spheres) found in a region of P18 box catalogue (black and red points) with number density $3\times10^{-3} h^{3}$Mpc$^{-3}$. Points that define the voids (i.e. those lying in their surface) are highlighted with a red circle. The volume of the box is $80^{3} h^{-3}$Mpc$^{3}$.}
    \label{p18voids}
\end{figure}

In Figure \ref{p18voids} the voids larger than $r>9h^{-1}$Mpc found in the P18 halo simulation box with number density $3\times10^{-3} h^{3}$Mpc$^{-3}$ are shown in a region of the box. It can be seen that each void is defined by four galaxies laying in its surface (some voids in Figure \ref{p18voids} have fewer than four galaxies due to being outside the plotted region) and we could ensure that voids do not overlap.

However, for the galaxy lightcones, one more step is required. This step involves considering the incompleteness in stellar mass, which causes many spurious voids to appear. These spurious voids correspond to regions in the catalogue with very low completeness, where galaxies could not be detected properly. Therefore, these false voids must be eliminated using an angular mask, such as the \texttt{Healpix} map \citep{1999astro.ph..5275G, 2005AJ....129.2562B, 10.1111/j.1365-2966.2008.13296.x},  characteristic of the redshift survey. 

The procedure for removing these false voids is as follows: first, we generated points uniformly distributed within the volume of the void. Next, we projected these points in the angular plane, calculated the completeness of each point, and averaged the completeness of all points within the void to obtain an approximation of the completeness of the void. If this completeness is equal to or greater than 0.9, we labelled that void as a true void. Otherwise, we removed the false void.

Additionally, voids whose centres or part of their volume are outside the lightcone in the radial direction must also be removed, as the procedure explained above considers only the angular plane; however, voids may extend beyond the sample in the radial direction, so this extra check is necessary.

\begin{table}[bth!]
\caption{Abundance of voids larger than $r$ from SDSS and Uchuu-SDSS voids.}
\centering
\begin{tabular}{ccc}
\toprule
r & SDSS & Uchuu-SDSS \\
\midrule
10 & 797 $\pm$ 12 & 792.2 $\pm$ 2.1 \\
11 & 550 $\pm$ 15 & 550 $\pm$ 3\\
12 & 381  $\pm$ 13 & 366.3 $\pm$ 2.3\\
13 & 248 $\pm$ 13 & 233.4 $\pm$ 2.2\\
14 & 150 $\pm$ 11 & 140.6 $\pm$ 1.9\\
15 & 85 $\pm$ 8 & 78.9 $\pm$  1.4\\
16 & 51 $\pm$ 7 & 41.4 $\pm$ 1.2\\
17 & 27 $\pm$ 5 & 21.2 $\pm$ 0.8\\
18 & 15 $\pm$ 3 & 9.3 $\pm$ 0.5\\
19 & 7.0 $\pm$ 1.7 & 3.4 $\pm$ 0.3\\
20 & 1.0 $\pm$ 1.0 & 0.97 $\pm$ 0.18\\
\bottomrule
\end{tabular}
\tablefoot{Abundance of voids larger than $r$, $N_{v}(r)$, obtained from voids found in the distribution of SDSS (second column) and Uchuu-SDSS (third column) galaxies with $M_{r}<-20.5$ and $0.02<z<0.132$.}
\label{nbarra_tabla}
\end{table}

\begin{figure}[bth!]
    \centering
    \includegraphics[width=.5\textwidth]{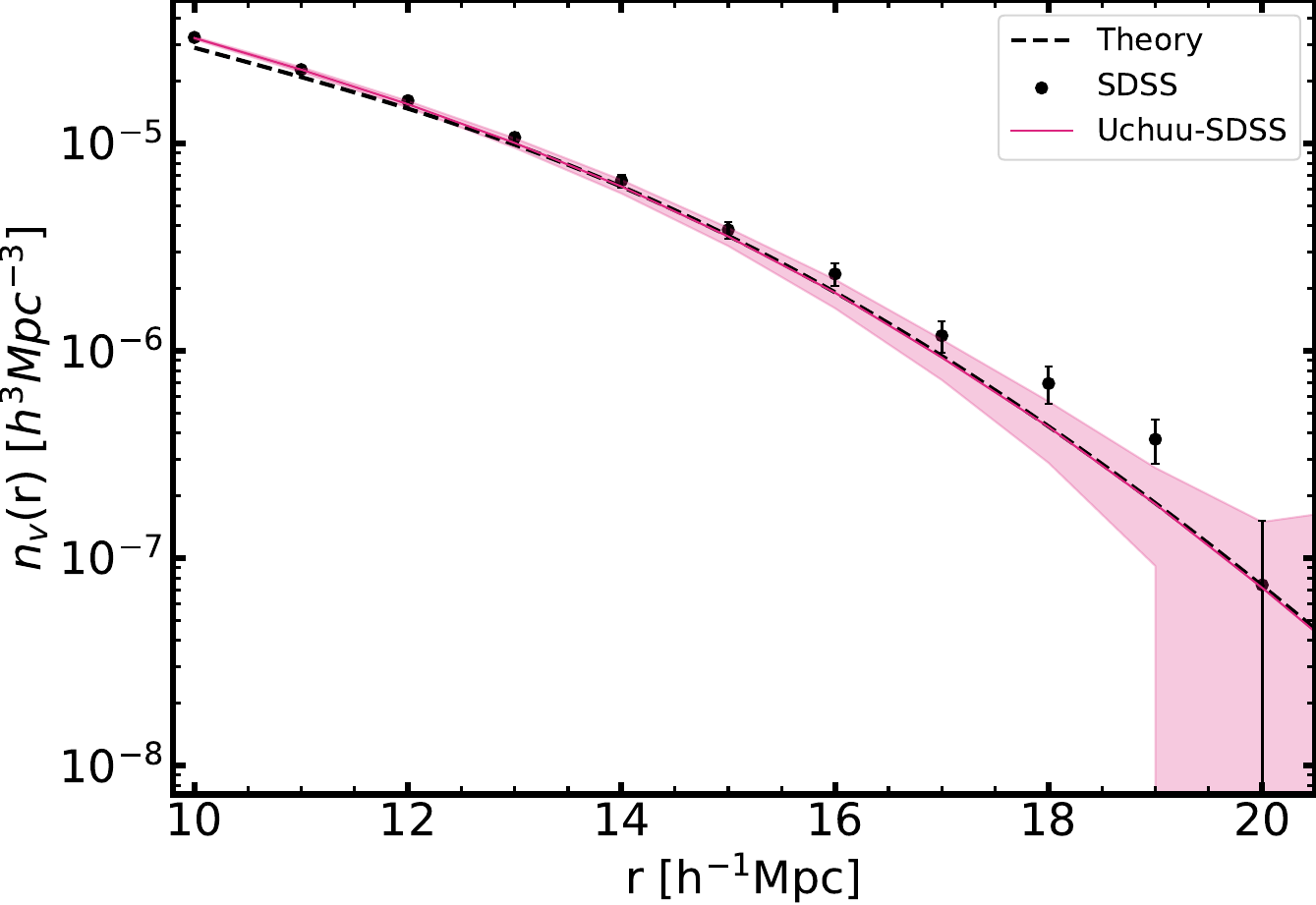}
    \caption{Number density of voids larger than $r$, $n_{v}(r)_{c}$, obtained for voids found in the distribution of SDSS (black points) and Uchuu-SDSS galaxies (pink solid line) with $M_{r}<-20.5$ and $0.02<z<0.132$ and predicted by the theoretical framework described in Section \ref{sec2} with Planck 2015 parameters (black dashed line). Shaded pink region delimits the standard deviation ($1\sigma$) of the 32 Uchuu-SDSS lightcones.}
    \label{nbarra_SDSS}
\end{figure}

In Figure \ref{nbarra_SDSS}, the number density of voids larger than $r$, $n_{v}(r)$, obtained for voids found in the sample of SDSS and Uchuu-SDSS (the mean of the 32 lightcones) considered in this work (galaxies with $M_{r}<-20.5$ and $0.02<z<0.132$) are shown (in Table \ref{nbarra_tabla} the values are multiplied by the volume). It can be checked that Uchuu-SDSS statistics are compatible with observations for all radius bins, and both are compatible with the values predicted by the theory.

The volume used in order to calculate the number density of voids larger than $r$ in Figure \ref{nbarra_SDSS} is:

\begin{equation}
    V(r) = \frac{N_{v}(r)(Uchuu-SDSS)}{n_{v}(r)(Uchuu)}
,\end{equation}
where $N_{v}(r)(Uchuu-SDSS)$ is the number of voids larger than $r$ found in Uchuu-SDSS lightcones and $n_{v}(r)(Uchuu)$ is the number density of voids larger than $r$ found in Uchuu galaxy box. Although this quantity has units of volume, it is not a real volume. However, this quantity allow us to convert lightcone statistics into number densities. This volume takes into account all the effects that can impact the volume that a real void can occupy, such as the completeness (amongst others).

An important remark about Figure \ref{nbarra_SDSS} is that if we want to compare the observed number density of voids larger than $r$ obtained from SDSS (or Uchuu-SDSS) with that given by the theoretical framework presented in this work, we must take into account that the former suffers from incompleteness, as well as other effects such as border effects, while the latter does not. Therefore, we have to transform SDSS (and Uchuu-SDSS) void statistics as if it did not suffer from these effects. One way of doing this is using Uchuu galaxy box. Then, SDSS and Uchuu-SDSS number density of voids larger than $r$ must be transformed as:

\begin{equation}
    n_{v}(r) \rightarrow n_{v}(r)(Uchuu)\frac{N_{v}(r)(SDSS)}{N_{v}(r)(Uchuu-SDSS)}
,\end{equation}
where $N_{v}(r)(Uchuu-SDSS)$ is the number of voids larger than $r$ found in Uchuu-SDSS lightcones, while $N_{v}(r)(Uchuu)$ is the same but found in Uchuu galaxy box.

\section{Void statistics theoretical framework }\label{sec2}
The void size function of voids was predicted for the first time \cite{10.1111/j.1365-2966.2004.07661.x}, where the authors used the same excursion-set approach used for the mass function of dark matter halos \cite{1974ApJ...187..425P}. This model is also based on the definition of voids as under-dense, non-overlapping spheres, which have gone through shell crossing (see \cite{1974ApJ...187..425P} for more details). However, this model predicts that the fraction of the volume occupied by voids can be larger than the total volume of the universe. This was fixed by \cite{10.1093/mnras/stt1169}, where they introduce the volume-conserving model (Vdn model), where the total volume occupied by cosmic voids is conserved in the transition to the non-linear regime. 

As demonstrated in \cite{2019MNRAS.488.5075R}, the Vdn model is able to  accurately predict the measured void size function of unbiased tracers, provided that the void catalogue is approximately cleaned from spurious voids and that the void radii are re-scaled to a fixed density threshold.  This means that the developed void finder has to find voids with these requirements. However, in many of the works where this model is used \citep{2019MNRAS.488.3526C, Contarini_2022, Contarini_2023, Contarini_2024}  a void finder based on Voronoi tesselations has been employed, which does not require voids to be spherical; then,  a cleaning algorithm was applied to remove small voids and transform the remaining voids, so that they are non-overlapping maximal spheres, with a specific density contrast threshold.

Additionally, the Vdn model predicts the number density of voids found in the distribution of dark matter particle field. In \cite{2019MNRAS.488.3526C}, an extension of this model is proposed by introducing a parametrisation of the Vdn model's non-linear under-density threshold as a function of the tracer effective bias, which has two free parameters that must be calibrated using simulations.

In this work, we propose a new model that only requires that voids be defined as maximal non-overlapping spheres. Thus, when using our void finder, we do not need to carry out a cleaning procedure. Additionally, the bias is completely determined by the model. 

In this section, we present the theoretical framework used in this work, which was present for the first time in \citep{art5}, to calculate the number density of voids larger than $r$. First, we show the equations in real space and then we explain how the equations change in redshift space. We compare the results provided by the theoretical framework with the results obtained with the four halo simulations to check its validity for different values of $\sigma_{8}$ and in real and redshift space. The last step is to compare the model with SDSS void statistics.

\subsection{Real space}
The main void statistics we will study in this work is the number density of voids larger than $r$, $\bar{n}_{v}(r)$\footnote[8]{Notice that we use a bar above $n_{v}(r)$ for quantities predicted by theoretical framework, and without bar for quantities obtained directly by simulations. Additionally, quantities in lowercase letters are in units of volume.}, which can be related to void probability function, $P_{0}(r)$ \citep{1979MNRAS.186..145W} using the (re-calibrated) expression from \cite{art5}:

\begin{equation}\label{numbervoids}
    \bar{n}_{v}(r) = \frac{\mu\mathcal{K}(r)}{V}e^{-\alpha\mathcal{K}(r)}
,\end{equation}
where $\mu$=0.588 and $\alpha$=1.671. These values have been calibrated using Uchuu simulation (see Appendix \ref{calibration_model}) and
\begin{equation}\label{eq}
    \mathcal{K}(r) = \left[ -\frac{1}{3}\frac{dlnP_{0}(r)}{dlnr} \right]^{3} P_{0}(r) 
,\end{equation}
with
\begin{equation}
    V = \frac{4}{3}\pi r^{3}
.\end{equation}

In Eq. (\ref{numbervoids}), it is assumed that $\bar{n}_{v}(r)$ is an universal functional of $P_{0}(r)$, so that independently of the clustering process underlying $P_{0}(r)$, $\bar{n}_{v}(r)$ is related to it by an unique expression. However, it has been shown that for white noise, $\mu=0.68$. Thus, it is clear that this coefficient has a dependence on the clustering properties of the objects considered. However, we have found that for the simulations that we have used the quoted values of $\mu$ and $\alpha$ are valid. Therefore we shall use these values in all our considerations.

It is important to remark that Eq. (\ref{numbervoids}) is only valid for $\mathcal{K}(r)\leq$0.46. For $\mathcal{K}(r)>$ 0.46,  $\bar{n}_{v}(r)=0.313/V$. $\mathcal{K}(r)$, measures the rareness of the voids.
The expression of $P_{0}(r),$ which is the void probability function (VPF) is derived in Appendix \ref{computation} and in \cite{art5}, where it is shown that this expression arises from first principles. Additionally, in Appendix \ref{validation_model}, we show that the VPF successfully reproduces the VPF of the four simulation boxes used in this work. 

Equations (\ref{Aeq}) and (\ref{beq}) were fit considering a cosmology with $\sigma_{8} = 0.9$, so these expressions are only valid for values of $\sigma_{8}$ around this value \citep{2008MNRAS.386.2181R}. Moreover, it was demonstrated that small voids do not carry any cosmological information, so in this work, we  only consider voids with $r>10h^{-1}$Mpc.
From Eqs. (\ref{numbervoids}) and (\ref{eq}) we can see that if we know the VPF, we automatically know $\bar{n}_{v}(r)$; thus, we chose to first study the VPF and then calculate $\bar{n}_{v}(r)$ using Eq. (\ref{numbervoids}).

We can define $P_{0}(r)$ from a more general statistic, that is, $P_{n}(r)$: the probability that a sphere of radius, $r$, placed at random within the distribution, contains $n$ objects. If we assume a Poisson process, we can then write $P_{n}(r)$ as \citep{1954AJ.....59..268L}:

\begin{equation}
    P_{n}(r) = \int_{0}^{\infty} P(u)\frac{u^{n}}{n!}e^{-u}du
    \label{Pn}
,\end{equation}
where $P(u)$ is the probability distribution for the integral of the probability density, $u$, within a randomly placed sphere. It is important to note that the dependence of $P_{n}(r)$ with $r$ is through $u(r)$ (see Eqs. (\ref{a2}) and (\ref{delf})).

An explicit computation of each term in $P_{n}(r)$ can be consulted in Appendix \ref{computation}. There, it can be seen that $P_{n}(r)$ depends on $\sigma_{8}$ (Eq. (\ref{sigmaeq})), $\Gamma=\Omega_{\rm c}h$ (Eqs. (\ref{Aeq}), (\ref{Beq}), and (\ref{Ceq})), and $\Omega_{\rm m}$ (Eqs. (\ref{sigma8redshift}) and (\ref{VELeq})).  However, the dependence on $\Omega_{\rm m}$ is small.

We can see that so far our theoretical framework depends on four cosmological parameters: $\sigma_{8}$, $\Omega_{\rm c}$, H$_{0}$, and $\Omega_{m}$. We note that the latter does not depend on $\Omega_{\Lambda}$, as we impose a flat $\Lambda$CDM model. Also,  the dependence of the theoretical framework with $\Omega_{\rm c}$ and H$_{0}$ is only through $\Gamma=\Omega_{\rm c}h$. Additionally, the dependence on $\Omega_{\rm m}$ is very weak, as it only enters through D(a). However, we can decrease the number of free parameters by relating $\Omega_{m}$ and $\Omega_{b}=\Omega_{m}-\Omega_{\rm c}$, where $\Omega_{b}$ is the baryonic matter density parameter. From Planck 2018 \citep{planck2018} we get $\alpha\equiv\Omega_{b}/\Omega_{m}=0.157$. Therefore, $\Omega_{\rm c}=(1-\alpha)\Omega_{m}$ and $\Omega_{b}=\alpha\Omega_{m}$, so our final parameters are $\sigma_{8}$, $\Omega_{m}$ and $H_{0}$.

Before using the theoretical framework to constrain $\sigma_{8}$, $\Omega_{\rm m}$, and H$_{0}$ in the SDSS survey, it is important to check if the theoretical equations are accurate and have enough precision to recover the real values of $\sigma_{8}$, $\Omega_{\rm m}$, and H$_{0}$ from simulations with different values of the parameters. To do this, we used the halo simulation boxes that we have presented in Section \ref{mocks}, which have different values of $\sigma_{8}$.

\begin{figure}[h!]
    \centering
    \includegraphics[width=.5\textwidth]{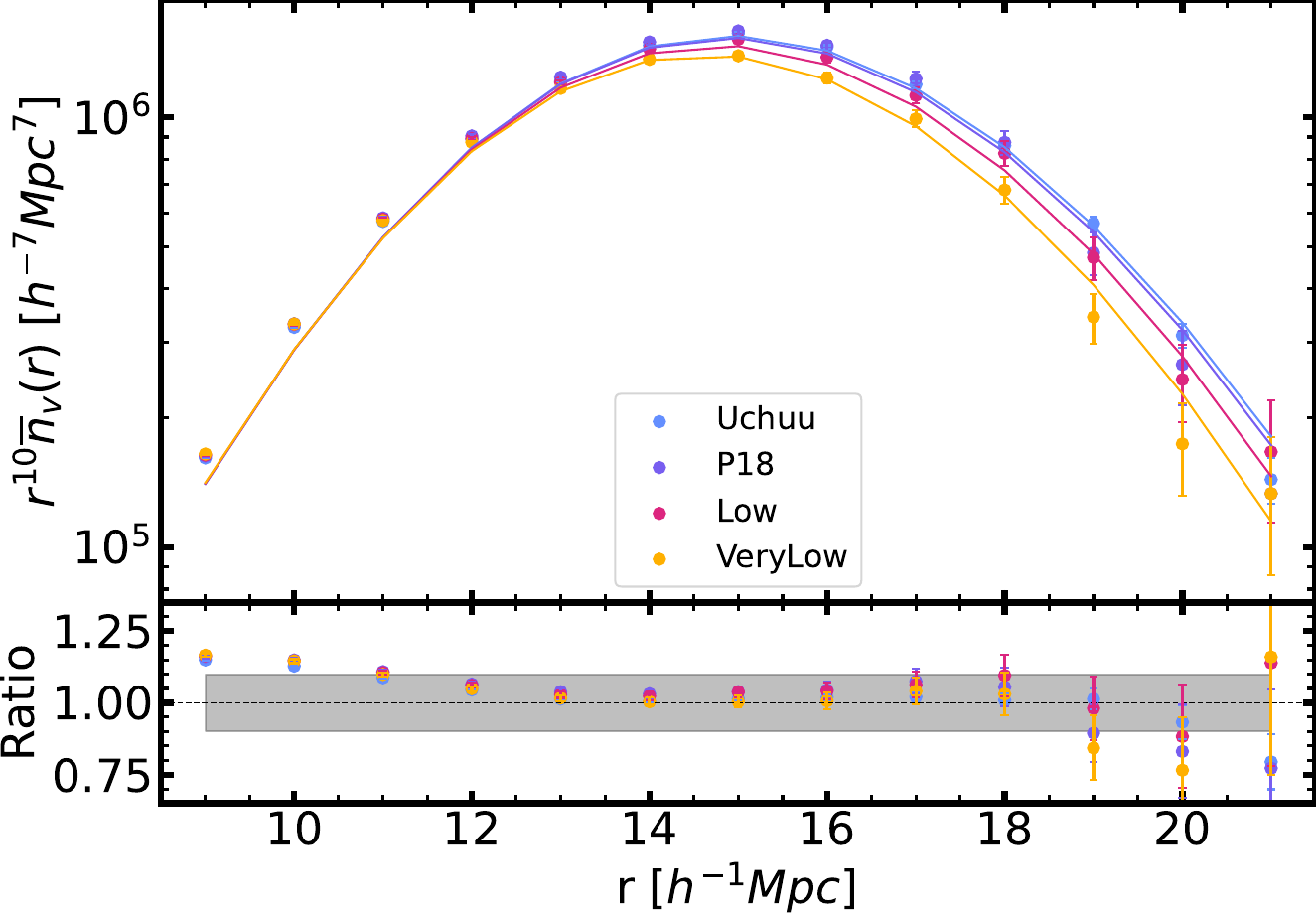}
    \caption{ Number density of voids larger than $r$ in real space is shown for the theoretical framework (lines) and simulations (dots), shown in the left panel. The ratio between simulations and theoretical framework is shown in the bottom panels for the Uchuu, P18, Low, and VeryLow box catalogues with a number density of $\bar{n}=3\times10^{-3} h^{3}$Mpc$^{-3}$. The shaded region in the bottom panel  delimits the region between 0.9 and 1.10 for the ratio.}
    \label{nvoids_halos}
\end{figure}

The upper panel of Figure \ref{nvoids_halos} gives the number density of voids larger than $r$ predicted by the theoretical framework (continuous lines) and obtained by simulations (points) for the four halo simulation boxes in real space can be seen. In the bottom panel of the same figure, the ratio between simulations and theoretical framework is shown. Again, the agreement between the theoretical framework and the simulation values is good, especially for $r$ between 12 and 18 $h^{-1}$Mpc.

\subsection{Redshift space}
The theoretical framework developed above is only valid for voids in real space. However, if we want to constrain $\sigma_{8}$ using surveys such as SDSS, this surveys provide galaxy positions in redshift space. In this space, the peculiar velocity of the galaxies is added to the velocity expansion of the Universe. This generates some distortions which result in elongated structures known as Fingers of God (see \cite{Hamilton_1998} for more details).
If we want to calculate $\bar{n}_{v}(r)$ and $P_{n}(r)$ in redshift space instead of real space, all we have to do is adjust $r$ via:

\begin{equation}
    r = r^{*}\times\{1+gVEL[\delta]\}^{-1/3}
    \label{rredshift}
,\end{equation}
where $r$ is the radius of a sphere in real space and $r^{*}$ is the radius of that sphere in redshift space. Additionally, it has been found comparing the average outflow around the relevant voids with that given by the spherical expansion model that the value of $g$ is around 0.85. This is also the value that provides the best agreement with simulations. 

VEL($\delta$) is defined so that the peculiar velocity, $V$, of mass element at distance $r$ from the centre of a spherical mass concentration (or defect) enclosing actual density contrast $\delta$ is given by
\begin{equation}
    V=HrVEL(\delta)
    \label{VELeq}
,\end{equation}
where $H$ is the Hubble constant at the time being considered. In \cite{Betancort-Rijo_2006}, it was shown that
\begin{equation}
    VEL(\delta)=-\frac{1}{3}\frac{dlnD(a)}{dlna}\frac{DELK(\delta)}{1+\delta}\left(\frac{d}{d\delta}DELK(\delta)\right)^{-1}
    \label{vel}
,\end{equation}
where $D(a)$ is the growth factor as a function of the expansion factor, $a$, and DELK($\delta$) is the inverse function of DELT($\delta_{l}$)  and next \cite[see][]{1996MNRAS.282..347M, 10.1046/j.1365-8711.2002.04950.x}:
\begin{equation}
\begin{aligned}
    DELK(\delta)&=\frac{\delta_{c}}{1.68647}\Bigg(1.68647-\frac{1.35}{(1+\delta)^{2/3}}-\frac{1.12431}{(1+\delta)^{1/2}}+\\&+\frac{0.78785}{(1+\delta)^{0.58661}}\Bigg),
\end{aligned}
\end{equation}
where $\delta_{c}$ is the linear density contrast for spherical collapse model; for the concordance cosmology, the value at present is 1.676.

Finally, the expression of ${\displaystyle \frac{dlnD(a)}{dlna}}$ is:

\begin{equation}
    \frac{dlnD(a)}{dlna} \approx 1.06\left(\frac{(1+z)^{3}}{[(1+z)^{3}+\Omega_{\Lambda}/\Omega_{\rm m}]} \right)^{0.6}
.\end{equation}

It is important to note that in real space, $\Omega_{\rm m}$ appears in the equations only through D(a), as per Eq. (\ref{sigmaeq})). However, we can see that in redshift space $\Omega_{\rm m}$ enters the theoretical framework also through VEL($\delta$) (see Eq. (\ref{vel})). Therefore, we would expect that the dependency of the $\bar{n}_{v}(r)$ on $\Omega_{\rm m}$ is stronger in redshift than in real space.

\begin{figure}[bth!]
    \centering
    \includegraphics[width=.5\textwidth]{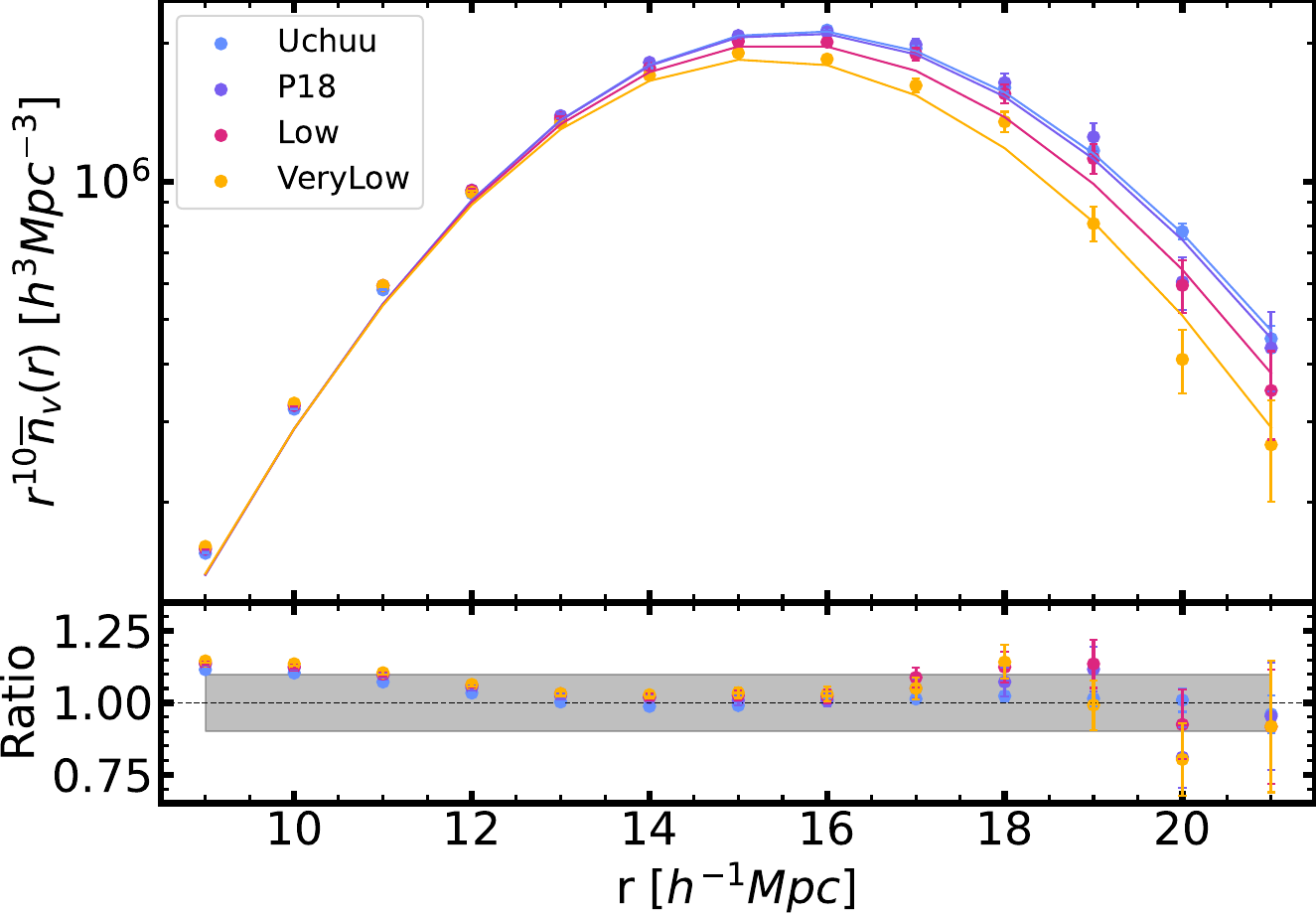}
    \caption{Number density of voids larger than $r$ in redshift space is shown for the theoretical framework (lines) and simulations (dots) in the left panel, while the ratio between simulations and theoretical framework is shown in the bottom panels for the Uchuu, P18, Low, and VeryLow box catalogues with a number density of $\bar{n}=3\times10^{-3} h^{3}$Mpc$^{-3}$. The shaded region in the bottom panel indicates delimits the region between 0.9 and 1.10 for the ratio.}
    \label{nvoids_halos_redshift}
\end{figure}

In Figure \ref{nvoids_halos_redshift}, the number density of voids larger than $r$ in redshift space is shown for the four halo simulation boxes. The agreement is still good, so we can conclude that the theoretical framework works in redshift space as well.
It is important to note that the values of $\bar{n}_{v}(r)$ for the three small box catalogues (each with a different value of $\sigma_{8}$) are very similar for small values of $r$ ($r<13h^{-1}$Mpc). The differences between the three simulations begin to become important for $r>16h^{-1}$Mpc approximately. We thus considered  only these voids  to constrain the cosmological parameters in the next section.

In Figure \ref{nbarra_SDSS} we have already shown the abundance of voids larger than $r$ for SDSS and Uchuu-SDSS. The values given by the theoretical framework are shown in the same figure with a continuous black line. We can see that the agreement between the theoretical framework and Uchuu-SDSS is good for $r>14h^{-1}$Mpc, namely, they are compatible within 1$\sigma$.

To sum up these findings, in this section we have checked that the theoretical framework successfully predicts the number density of voids larger than $r$ for the four halo simulation boxes in real and redshift space. We have also shown that the prediction from Uchuu-SDSS voids is compatible with SDSS voids statistics within $1\sigma$ and with the prediction of the theoretical framework. Therefore, we are ready to constrain $\sigma_{8}$ and $\Omega_{\rm m}$ and H$_{0}$ from the SDSS galaxy sample we have chosen using the theoretical framework in redshift space.

\section{Bayesian analysis for cosmological parameters inference}\label{likelihoodtest}
To infer $\sigma_{8}$, $\Omega_{\rm m}$ and H$_{0}$ from the number density of voids larger than $r$, we used a Bayesian analysis and Markov chain Monte Carlo (MCMC) technique to sample the posterior distribution of the considered parameter sets, $\Theta$:

\begin{equation}
    \mathcal{P}(\Theta|\mathcal{D}) \propto \mathcal{L}(\mathcal{D}|\Theta)p(\Theta)
    \label{bayes}
,\end{equation}

where $p(\Theta)$ is the prior distribution, and $\mathcal{L}(\mathcal{D} \mid \Theta)$ is the likelihood, calculated as

\begin{equation}
    \mathcal{L}(\mathcal{D}|\Theta) = \prod_{i=1}^{6}\frac{1}{\sigma_{i}}exp\left(-\frac{\left(N_{v,i}(\mathcal{D})-\bar{N}_{v,i}(\Theta)\right)^{2}}{2\sigma_{i}^{2}}\right)
    \label{likelihood}
.\end{equation}

$N_{v,i}(\mathcal{D})$ is the number of voids within $i$th bin from simulations or data, $\bar{N}_{i}(\Theta)$ is calculated from Eq. (\ref{numbervoids}) subtracting the value in the $(i+1)$th bin (or $r+\Delta r$), and $\sigma_{i}$ is the rms of $N(r_{i})$:
\begin{equation}
    \sigma_{i}=rms(\bar{N}_{v,i}(\Theta)) = \sqrt{\bar{N}_{i}(\Theta)/N_{r}}
,\end{equation}
where $N_{r}$ represents the number of realisations for each simulation. For the simulation boxes and SDSS, $N_{r}$ is equal to 1, indicating a single realisation for each set of cosmological parameters and one SDSS survey. However, in the case of Uchuu-SDSS, $N_{r}$ is 32, as we have 32 lightcones of this particular type.

In this work, we use voids larger than $r=16h^{-1}$Mpc to constrain the cosmological parameters of interest, so in Eq. (\ref{likelihood}) we consider the radius bins starting from $r=16h^{-1}$Mpc to $r=21h^{-1}$Mpc (i.e. 6 bins with width $\Delta r$= 1$h^{-1}$Mpc).

With Eq. (\ref{bayes}) the best estimate of the parameters $\Theta$ may be obtained by maximising $\mathcal{P}(\Theta|\mathcal{D})$ with respect to the parameters $\Theta$. To do so, we assigned to all the parameters ($\sigma_{8}$, $\Omega_{\rm m}$, and H$_{0}$) wide enough priors (see first column of Table \ref{parameter_ranges}). 

\begin{table}[htb!]
\caption{Uniform priors used for the Bayesian analysis in this work.}
    \centering
    \begin{tabular}{ccc}
    \toprule
          &Priors A&  Priors B\\
         \midrule
         $\sigma_{8}$ &[0.5-2.5]&  [0.5-2.2] \\
         $\Omega_{\rm m}$ &[0.15-0.9]&  [0.15-0.686] \\
         H$_{0}$ &[55-91]&  [64-82] \\
         \bottomrule
    \end{tabular}
    \tablefoot{Limits of the uniform flat priors used in order to constrain these parameters considering only the void statistics developed in this work (second column), and in order to compare our marginalised posteriors with DESY3 (second column) and with KiDS-1000 and KiDS-1000+DESY3 (third column)  in Figure \ref{s8SDSScontour}.}
    \label{parameter_ranges}
\end{table}

To infer $\sigma_{8}$, $\Omega_{\rm m}$, and H$_{0}$ from SDSS and Uchuu-SDSS voids, we use MCMC sampler from \texttt{Cobaya} \citep{2019ascl.soft10019T, Torrado_2021}. We assess chain convergence using the Gelman-Rubin test \citep{10.1214/ss/1177011136}. Once a tolerance of 0.01 is achieved, we consider the chains to have converged. Moreover, we discarded the first 30$\%$ of each chain as burn-in and used the mean along with the 68\% two-tail equal-area confidence limit to represent the best-fit value and its uncertainty.

Additionally, we may want to combine our results with other works. In this work, we consider Planck 2018 \citep{planck2018} and two weak lensing results: KiDS-1000 and DESY3  \citep{2023OJAp....6E..36D}.  To do so, we used a brand new code called \texttt{CombineHarvesterFlow} \footnote{\url{https://github.com/pltaylor16/CombineHarvesterFlow}}, which allowed us to efficiently sample the joint posterior of two non-covariant experiments with a large set of nuisance parameters \citep{taylor2024combineharvesterflow}. Specifically, \texttt{CombineHarvesterFlow} trains normalising flows on posterior samples to learn the marginal density of the shared parameters. Then, by weighting one chain by the density of the second flow, we can find the joint constraints.

The marginalised posteriors in the plane $\sigma_{8}$-H$_{0}$ (fixing $\Omega_{\rm m}$ in each case to its real value; see Table \ref{parametros_3}) obtained for the four halo simulation boxes can be seen in the Appendix \ref{appendixhaloes}. There, we show that we recover the values of $\sigma_{8}$ and H$_{0}$ of the simulations within $1\sigma$ ($2\sigma$) in real (redshift) space. Therefore, we can infer these cosmological parameters in Uchuu-SDSS lightcones.

\section{Cosmological constraints from SDSS survey}\label{SDSS_como_constraints}

In this section, we show the constraints obtained from SDSS redshift survey. Firstly, we show the constraints we obtain directly from the theoretical framework developed in this work and, finally, we combine our likelihood with the KIDS-1000 and DESY3 weak lensing results.

\subsection{SDSS voids-only constraints}

\begin{figure}[tbh!]
    \centering
    \includegraphics[width=.5\textwidth]{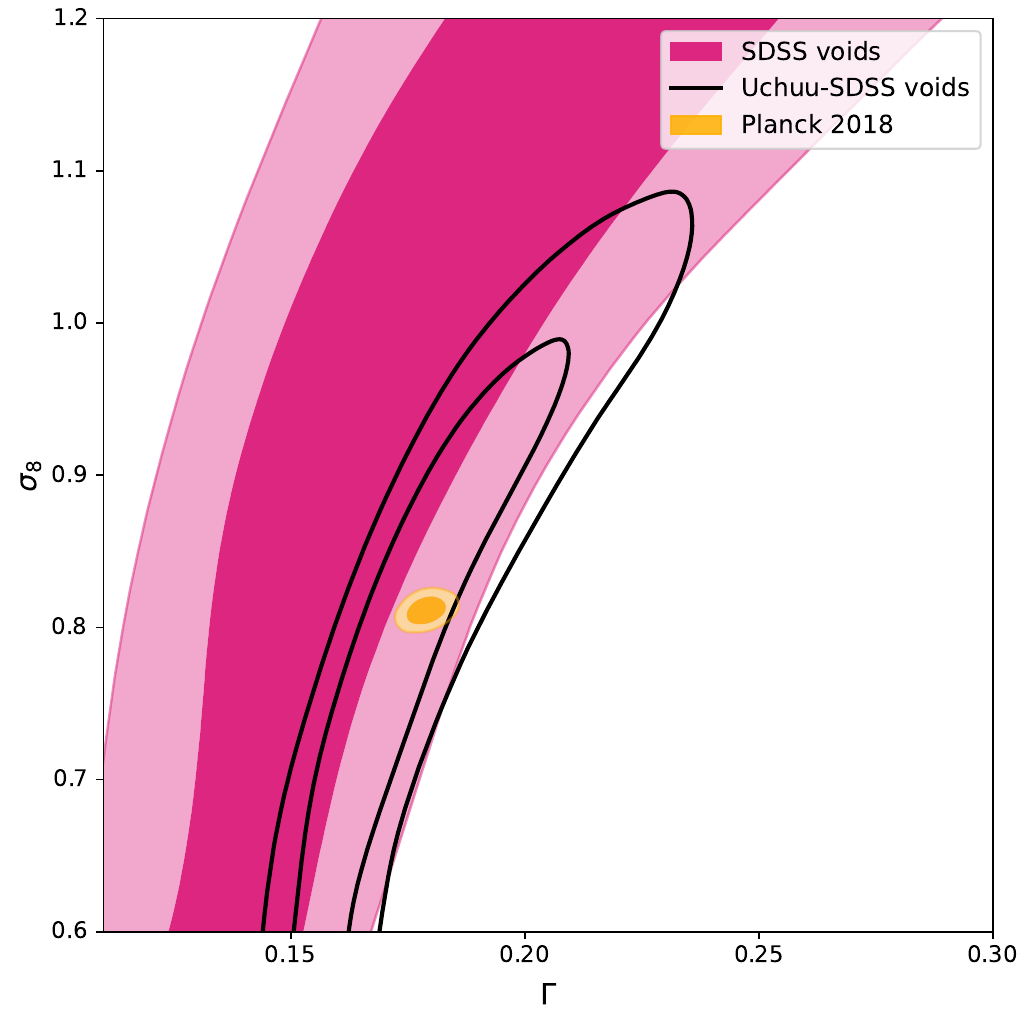}
    \caption{Marginalised posteriors in $\sigma_{8}-\Gamma$ plane obtained for Uchuu-SDSS voids and SDSS voids voids using the maximum likelihood test with Bayesian approach with the theoretical framework developed in this work. We also show the marginalised posteriors for Planck 2018. The contours indicate the 68$\%$ (1$\sigma$) and 95$\%$ (2$\sigma$) credible intervals.}
    \label{sigma8_gamma}
\end{figure}

The initial contour we examine lies within the plane showcasing the core parameters of our theoretical framework: the $\sigma_{8}$-$\Gamma$ plane, as depicted in Figure \ref{sigma8_gamma}. In this figure, we have also included the marginalised posteriors for Uchuu-SDSS and Planck 2018.

From Figure \ref{sigma8_gamma}, it can be seen that the marginalised posterior region for Planck 2018 is entirely encapsulated within the SDSS contour at the region of 2$\sigma$, indicating that the constraints from both samples are statistically compatible in this limit. Therefore, we won't combine SDSS with Planck 2018: the combined marginalised posteriors will be similar to the marginalised posteriors from Planck 2018.

\begin{figure*}[tbh!]
    \centering
    \includegraphics[width=.49\textwidth]{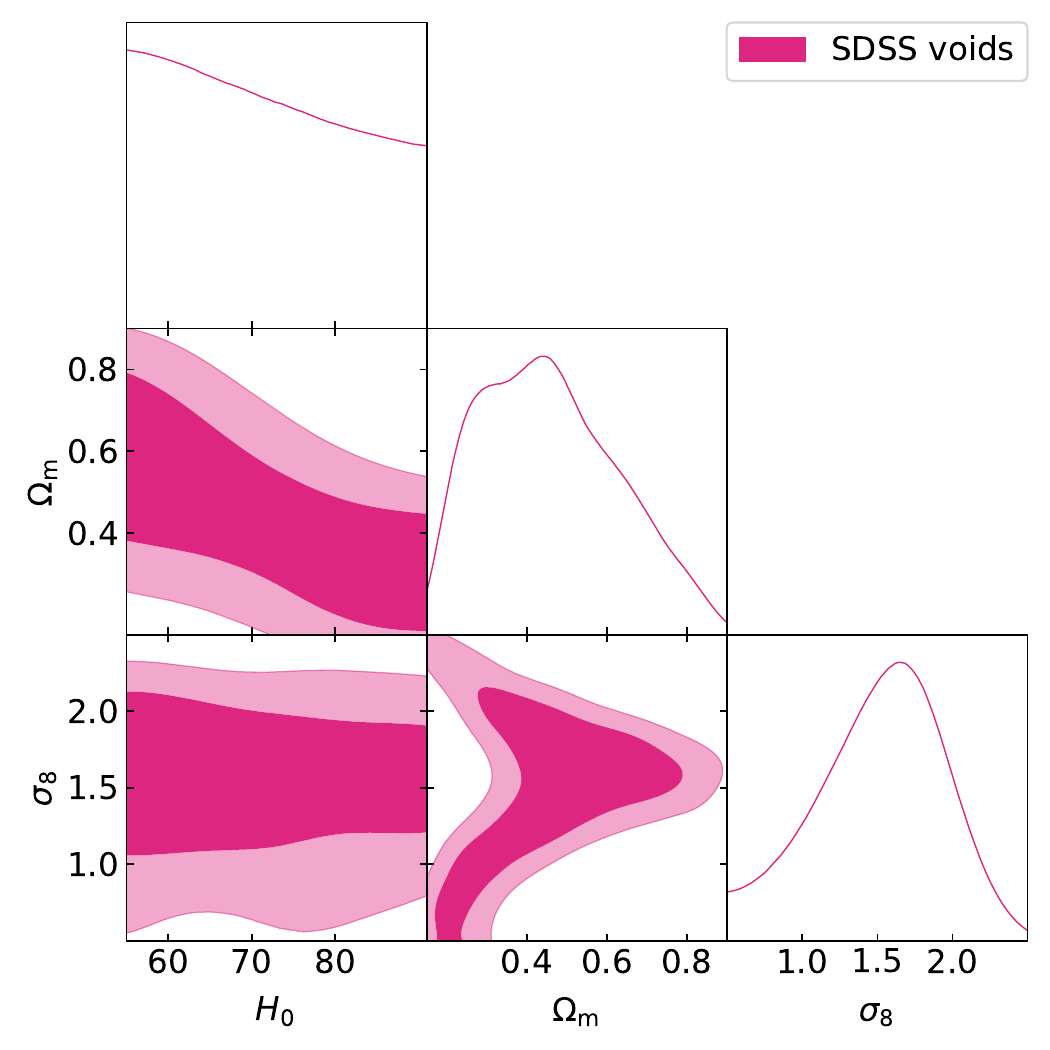}
    \includegraphics[width=.49\textwidth]{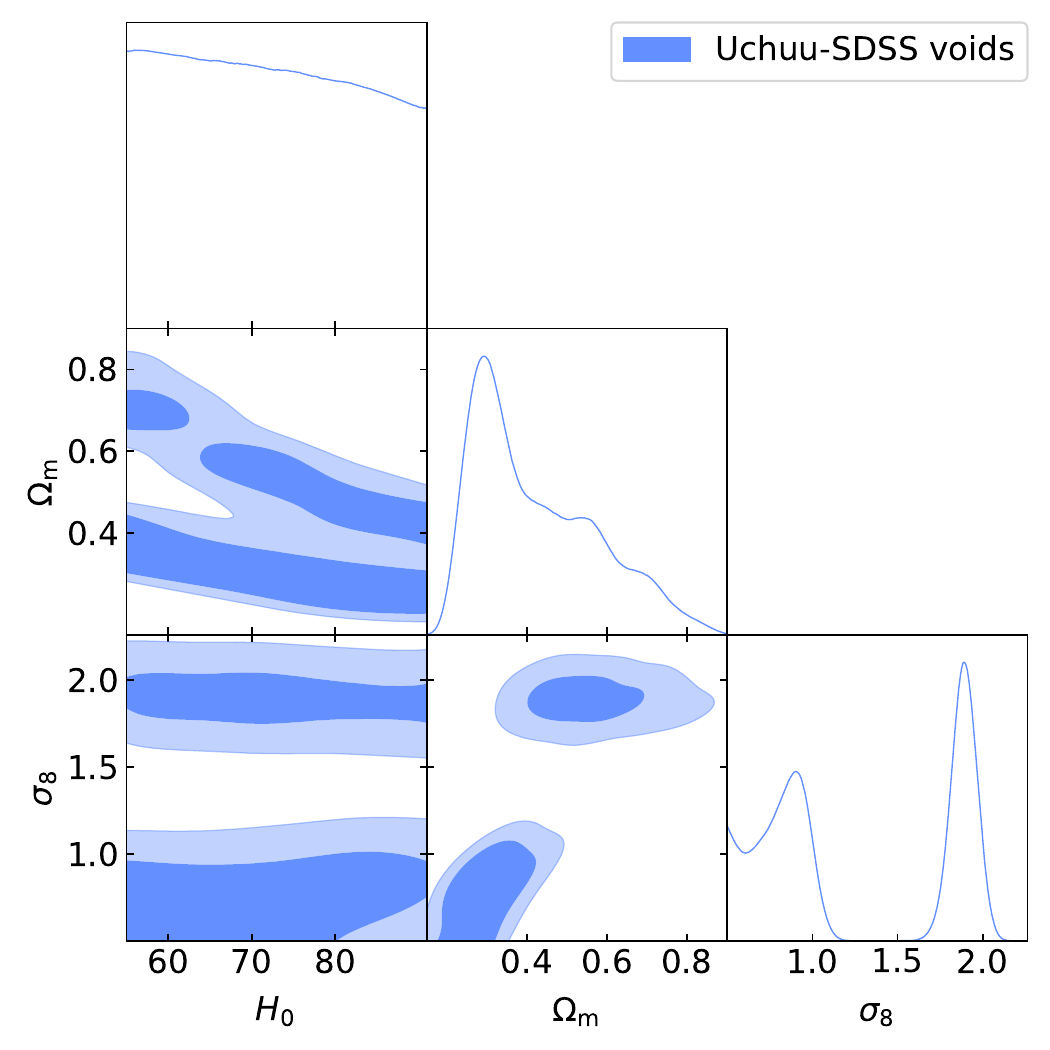}
    \caption{Marginalised posteriors in $\Omega_{\rm m}$-H$_{0}$, $\sigma_{8}$-H$_{0}$, and $\sigma_{8}-\Omega_{\rm m}$ planes obtained for SDSS redshift survey (left) and Uchuu-SDSS lightcones (right) using the maximum likelihood test with Bayesian approach with the theoretical framework developed in this work. The contours indicate the 68$\%$ (1$\sigma$) and 95$\%$ (2$\sigma$) credible intervals.}
    \label{SDSScontours}
\end{figure*}

In the left part of Figure \ref{SDSScontours} we can observe the marginalised posteriors obtained in the rest of the planes: $\Omega_{\rm m}$-H$_{0}$, $\sigma_{8}$-H$_{0}$, and $\sigma_{8}-\Omega_{\rm m}$. In the right part of the same figure, we can observe the marginalised posteriors obtained from Uchuu-SDSS.

In the first and second columns of Table \ref{constraints_planck}, we can observe the constrained values directly obtained from our theoretical framework of $\sigma_{8}$, $\Omega_{\rm m}$, H$_{0}$, $\Gamma$, and S$_{8}$ from Uchuu-SDSS lightcones and SDSS survey, respectively. From this table, it can be seen that the uncertainties of $\sigma_{8}$ and $\Gamma$ are much larger for SDSS than for Uchuu-SDSS because of the huge difference in the volumes between them. However, this is not the case for $\Omega_{\rm m}$ and H$_{0}$. This is because $\sigma_{8}$ and $\Gamma$ are the fundamental parameters of the theoretical framework (and S$_{8}$ depends strongly on $\sigma_{8}$). With these large errors, our constraints from SDSS are compatible with Planck's within 1$\sigma$.

\begin{table}[tbh!]
\caption{Constraints of the cosmological prameters from Uchuu-SDSS and SDSS voids compared to Planck 2018.}
    \centering
    \renewcommand{\arraystretch}{1.5} 
    \begin{tabular}{cccc}
    \toprule
            &Uchuu-SDSS voids&SDSS voids &Planck\\
            \midrule
          $\sigma_{8}$  &1.296$^{+0.621}_{-0.592}$&1.520$^{+0.416}_{-0.441}$&0.8111$\pm$0.0060\\
          $\Omega_{\rm m}$  &0.419$^{+0.172}_{-0.152}$&0.459$^{+0.184}_{-0.184}$&0.3153$\pm$0.0073\\
          H$_{0}$    &75.35$^{+12.24}_{-11.85}$&71.63$^{+12.60}_{-11.70}$&67.36$\pm$0.54\\
          $\Gamma$    &0.2500$^{+0.096}_{-0.088}$&0.270$^{+0.094}_{-0.100}$&0.1772$\pm$0.0027\\
          S$_{8}$  &1.61$^{+1.05}_{-0.93}$&1.87$^{+0.59}_{-0.76}$&0.832$\pm$0.013\\
          \bottomrule
    \end{tabular}
    \tablefoot{Constraints of $\sigma_{8}$, $\Omega_{\rm m}$, H$_{0}$ (in units of kms$^{-1}$Mpc$^{-1}$), S$_{8}=\sigma_{8}\sqrt{\Omega_{\rm m}/0.3}$, and $\Gamma=\Omega_{\rm c}h$ for Uchuu-SDSS voids (second column) SDSS voids (third column) and Planck 2018 best-fit values \citep{planck2018} (fourth  column), with errors calculated as the 68$\%$ uncertainties.}
    \label{constraints_planck}
\end{table}

\subsection{SDSS voids + Weak lensing}
\begin{table*}[tbh!]
\caption{Cosmological constraints from SDSS voids compared to KiDS-1000, DESY3, SDSS voids combined with KiDS-1000, DESY3 and KiDS-1000+DESY3, and Planck 2018.}
    \centering
    \renewcommand{\arraystretch}{1.5} 
    \begin{tabular}{cccccccc}
    \toprule
           &SDSS voids & KiDS-1000    &DESY3&SDSS+KiDS &SDSS+DESY3 &SDSS+KiDS+DESY3&Planck\\
            \midrule
          $\sigma_{8}$ &1.520$^{+0.416}_{-0.441}$& 0.833$^{+0.133}_{-0.146}$   &0.816$^{+0.076}_{-0.065}$&0.859$^{+0.050}_{-0.051}$&0.881$^{+0.049}_{-0.047}$&0.858$^{+0.040}_{-0.040}$&0.8111$\pm$0.0060\\
          $\Omega_{\rm m}$ &0.459$^{+0.184}_{-0.184}$& 0.270$^{+0.056}_{-0.102}$   &0.297$^{+0.040}_{-0.060}$&0.247$^{+0.027}_{-0.026}$&0.256$^{+0.024}_{-0.023}$&0.257$^{+0.023}_{-0.020}$&0.3153$\pm$0.0073\\
          H$_{0}$   &71.63$^{+12.60}_{-11.70}$& 75.42$^{+1.70}_{-2.13}$ &73.83$\pm$4.98&74.20$^{+5.05}_{-5.18}$&74.72$^{+5.36}_{-5.56}$&74.17$^{+4.66}_{-4.66}$&67.36$\pm$0.54\\
          S$_{8}$ &1.87$^{+0.59}_{-0.76}$& 0.764$^{+0.031}_{-0.023}$   &0.802$^{+0.023}_{-0.019}$&0.775$^{+0.020}_{-0.019}$&0.811$^{+0.019}_{-0.017}$&0.794$^{+0.016}_{-0.016}$&0.832$\pm$0.013\\
        \bottomrule

    \end{tabular}
    \tablefoot{Constraints of $\sigma_{8}$, $\Omega_{\rm m}$, H$_{0}$ (in units of kms$^{-1}$Mpc$^{-1}$) and S$_{8}=\sigma_{8}\sqrt{\Omega_{\rm m}/0.3}$ for SDSS voids (second column), KiDS-1000 (third column), DESY3 (fourth column)  \citep{2023OJAp....6E..36D},  SDSS voids+KiDS-1000 (fifth column), SDSS+DESY3 (sixth column), SDSS+KiDS-1000+DESY3 (seventh column) and Planck 2018 \citep{planck2018} (last column), with errors calculated as the 68$\%$ uncertainties. 
    }
    \label{constraints_wl}
\end{table*}

Finally, we can combine our results with KiDS-1000 and DESY3. These results are presented in \cite{2023OJAp....6E..36D}. When combining our results with other studies, such as those mentioned above, caution must be taken, as the same range for all the parameters must be considered if we want to visually compare the marginalised posteriors obtained in each case and combine the chains using \texttt{CombineHarvesterFlow}. This allow us to make a fair comparison of our marginalised posteriors with those of weak lensing. However, as we have seen previously, the theoretical framework used in this work depends on the parameters $\sigma_{8}$, $\Omega_{\rm m}$, and H$_{0}$. In weak lensing studies, A$_{s}$ (DESY3) or S$_{8}$ (KiDS and KiDS-1000+DESY3) are sampled instead of $\sigma_{8}$, and $\omega_{c}$ and $\omega_{b}$ instead of $\Omega_{\rm m}$ (KiDS and KiDS-1000+DESY3). Therefore, we have to determine which priors on $\sigma_{8}$ and $\Omega_{\rm m}$ correspond to the aforementioned priors, ensuring compatibility across these parameters. Nevertheless, it is important to consider a significant limitation: some approximations from our theoretical framework are not valid for $\sigma_{8}$ values lower than 0.5, as mentioned earlier, nor for $\Omega_{\rm m}$ values lower than 0.15. Thus, we have added this additional condition in the priors. This may slightly affect the process of combining chains using \texttt{CombineHarvesterFlow}, but, as we will see later, the effect will be very small because of the relative orientation and size of SDSS voids and weak lensing marginalised posteriors.
The ranges we have used for our chains when combining with DESY3 and KiDS-1000 can be seen in the first and second columns, respectively, of Table \ref{parameter_ranges}.  We can see that the range of H$_{0}$ used in these works is very narrow. 

\begin{figure*}[tbh!]
    \centering
    \includegraphics[width=0.32\textwidth]{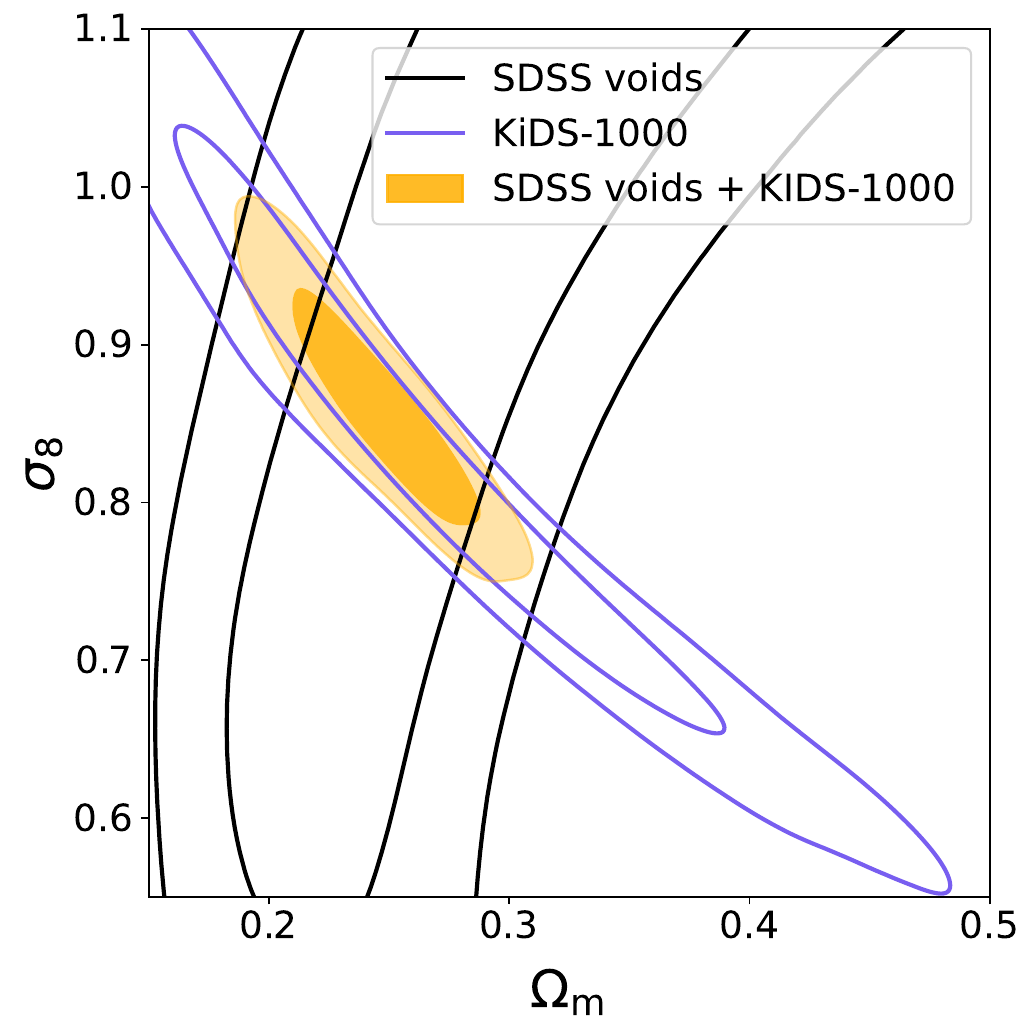}
    \includegraphics[width=0.32\textwidth]{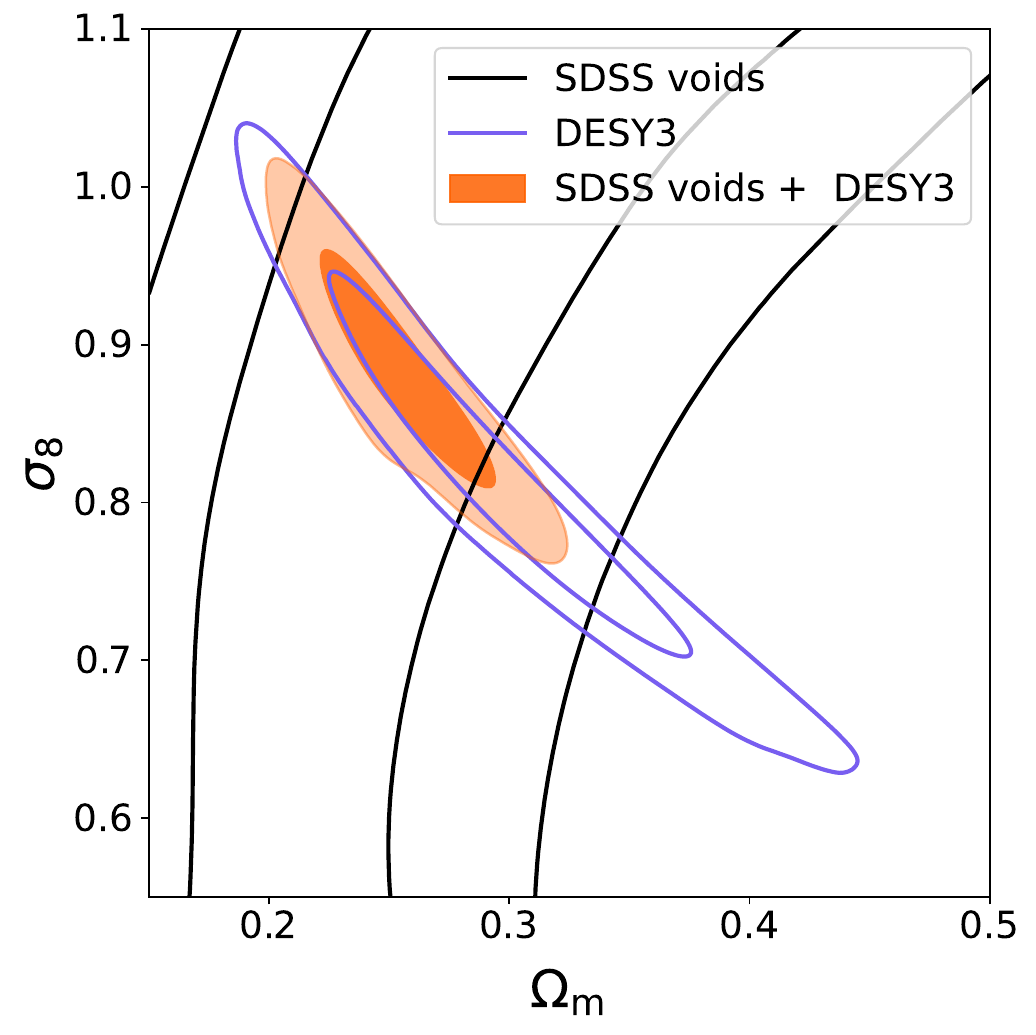}
    \includegraphics[width=0.32\textwidth]{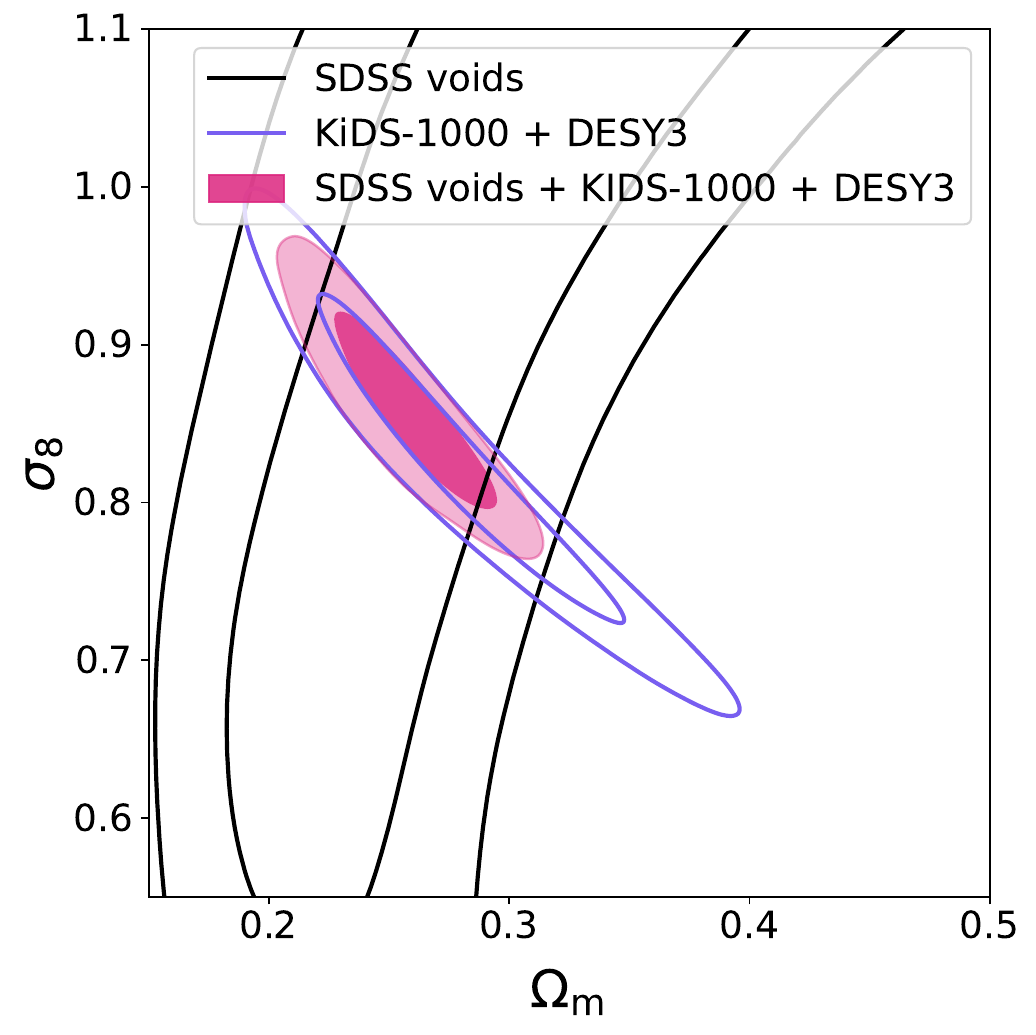}
    \caption{KiDS-1000, DESY3, and KiDS-1000+DESY3 contours, shown in the left, middle, and right panels, respectively. The SDSS voids contours are given in the  $\sigma_{8}-\Omega_{\rm m}$ plane, using the same parameter ranges to make a fair comparison, along with the combination of SDSS voids with each weak lensing work. In Table \ref{parameter_ranges}, the parameter ranges of each work are indicated. } 
    \label{s8SDSScontour}
\end{figure*}

The marginalised posteriors in the plane $\sigma_{8}-\Omega_{\rm m}$ using these ranges can be seen in Figure \ref{s8SDSScontour}. In this figure, the marginalised posteriors from SDSS voids with the same ranges as the different weak lensing works considered are shown, as well as each weak lensing work marginalised posteriors and the combination of SDSS voids with these three contours. We can see that SDSS voids contour is almost orthogonal to the three weak lensing contours (as expected, see \cite{Contarini_2023} for more details); thus, we can anticipate that our constrained values of $\sigma_{8}$ and $\Omega_{\rm m}$ will have smaller uncertainties than the values of the original work. 

The best-fit values of $\sigma_{8}$, $\Omega_{\rm m}$, H$_{0}$, and S$_{8}$ obtained from SDSS voids, KiDS-1000 and DESY3  can be seen in the first three columns of the Table \ref{constraints_wl}. The combination of SDSS voids with KiDS-1000, DESY3, and KiDS-1000+DESY3 can be seen in the fourth, fifth, and sixth columns, respectively, of the same table.\footnote{It is important to remark that \texttt{CombineHarvesterFlow} gives the inferred parameters obtained from the combination for two chains in two different ways: weighting SDSS voids chains or weighting weak lensing chains. The results we have presented correspond to weighting SDSS void chains, but we have checked that weighting weak lensing chains we obtain compatible results.} We can see that the effect of combining these three weak lensing works with SDSS voids is increasing the value of $\sigma_{8}$, as the best-fit value of SDSS voids is very high, which implies a decrease in the best-fit value of $\Omega_{\rm m}$ (from Figure \ref{s8SDSScontour} we can see that the combination of the three weak lensing works predicts a strong correlation between $\sigma_{8}$ and $\Omega_{\rm m}$, and that if one increases, the other decreases. This correlation is kept also when combining weak lensing works with SDSS voids).

We can also see from Table \ref{constraints_wl} that there is an increase in the uncertainties of H$_{0}$ when combining weak lensing with SDSS voids, which is caused by our huge uncertainties in this parameter. Specifically, the uncertainties when combining with SDSS void statistics with KiDS increase by a factor of 2-3 and a factor of 1.1 when combined with DESY3.  However, the uncertainties of the rest of the parameters are decreased by a factor 2-3, approximately, with respect to the original errors of each weak lensing work.

As a final remark, we can see from Table \ref{constraints_wl} that any value of H$_{0}$ or $\Omega_{\rm m}$ obtained when combining SDSS voids with the three weak lensing works are compatible with Planck 2018 within 1$\sigma$ (for $\Omega_{\rm m}$, they are within 2.3$\sigma$, approximately). We can also see that S$_{8}$ value from SDSS voids + DESY3 is compatible with Planck 2018 within 1$\sigma$, but SDSS voids + KiDS-1000 and SDSS voids + KiDS-1000+DESY3 are not (noting they are compatible with Planck 2018, considering the uncertainty as 1.8$\sigma$ and 1.6$\sigma$, respectively).

\section{Comparison with other works about voids}\label{comparison_sec}

In this section, we compare our constraints on $\sigma_{8}$, H$_{0}$, and S$_{8} = \sigma_{8}\sqrt{\Omega_{\rm m}/0.3}$ with those obtained in other studies that also utilise void statistics to constrain these parameters.

The first study to compare our results with is \cite{Contarini_2024}, where the values of S$_{8}$ and H$_{0}$ are constrained using the void size function (the abundance of voids with radius $r$) predicted by the excursion set theoretical framework \citep{1974ApJ...187..425P, 10.1111/j.1365-2966.2004.07661.x}, employing an extension of the widely used volume-conserving model \cite[Vdn model,][]{10.1093/mnras/stt1169}. The constraints they derived by combining void counts with void shapes \citep{Hamaus_2020} are $\sigma_{8} = 0.809^{+0.072}{-0.068}$, $\Omega_{\rm m} = 0.308^{+0.021}{-0.018}$, H$_{0} = 67.3^{+10.0}{-9.1}$, and S$_{8} = 0.813^{+0.093}_{-0.068}$. To constrain these cosmological parameters, they used data from the BOSS DR12 redshift survey \citep{2013AJ....145...10D}; specifically, the LOWZ and CMASS target selections. The catalogues are divided into two redshift bins: $0.2 < z \leq 0.45$ and $0.45 < z < 0.65$. This sample encompasses a physical volume approximately 60 times larger than the one used in this work. According to \cite{10.1093/mnras/stv2382}, the CMASS volume is 5.1 Gpc$^{3}$, and the LOWZ volume is 2.3 Gpc$^{3}$, resulting in a total volume of $2.3 \times 10^{9}$ $h^{-3}$Mpc$^{3}$. If we consider the scaling of errors with survey volume (as demonstrated in Appendix \ref{scaling_errors}) and apply this scaling to a redshift survey like BOSS DR12 using our theoretical framework, we can estimate how much the uncertainties would decrease. For $\Gamma$ (a parameter that was not constrained in \cite{Contarini_2024}), our errors would scale by a factor of $60^{-1/2.5} = 0.194$, making them five times smaller than those obtained in this work with the SDSS sample volume. For $\sigma_{8}$, the uncertainties would scale by $60^{-1/7} = 0.557$, reducing them by a factor of 1.79. This would result in uncertainties approximately 30\% larger than those reported in \cite{Contarini_2024}.  For S$_{8}$, the uncertainties would scale by $60^{-1/6} = 0.505$, reducing them by a factor of 1.98. This again implies uncertainties roughly 30\% larger than those in \cite{Contarini_2024}. However, for H$_{0}$ and especially $\Omega_{\rm m}$, there would be no significant improvement due to the limited scaling of these parameters with survey volume. As a result, the uncertainties for H$_{0}$ and $\Omega_{\rm m}$ would remain significantly larger than those obtained in \cite{Contarini_2024}.
It is important to note that the constraints provided in the aforementioned study are derived from the combination of the Vdn model and void shapes, whereas the constraints presented in this work are based solely on void statistics.

Another study to compare our results with is \cite{Sahl_n_2016}, where galaxy cluster and void abundances are combined using extreme-value statistics. This approach focuses on a large galaxy cluster and a void aligned with the Cold Spot (CS) in the CMB (referred to as the CS void) \citep{2016MNRAS.455.1246F}. Using this method, they obtained a constraint of $\sigma_{8} = 0.95 \pm 0.21$ for a flat $\Lambda$CDM universe. This result is consistent with values derived from the CMB as well as weak lensing studies.

\begin{figure*}[tbh!]
    \centering
    \includegraphics[width=.48\textwidth]{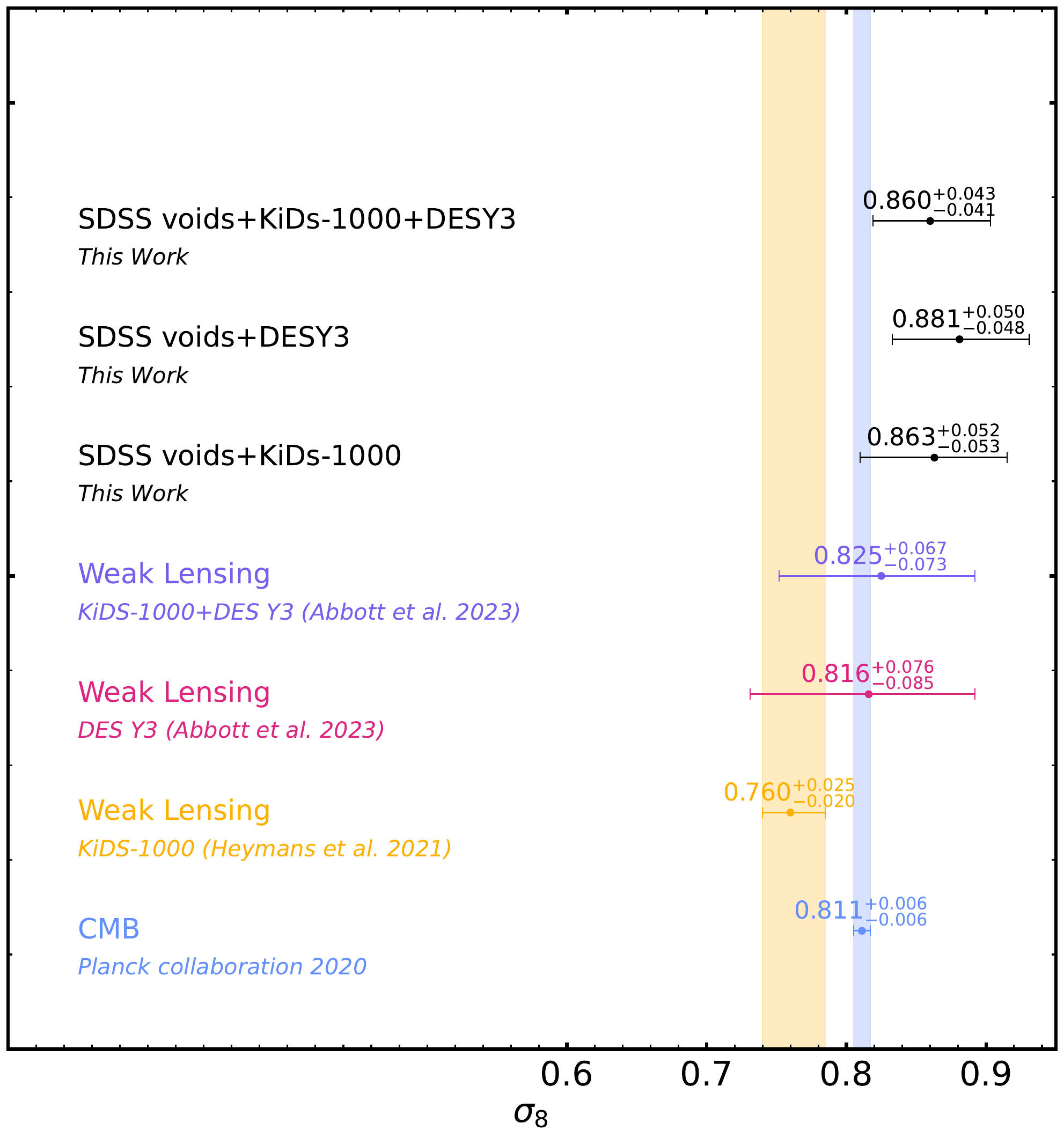}
    \includegraphics[width=.48\textwidth]{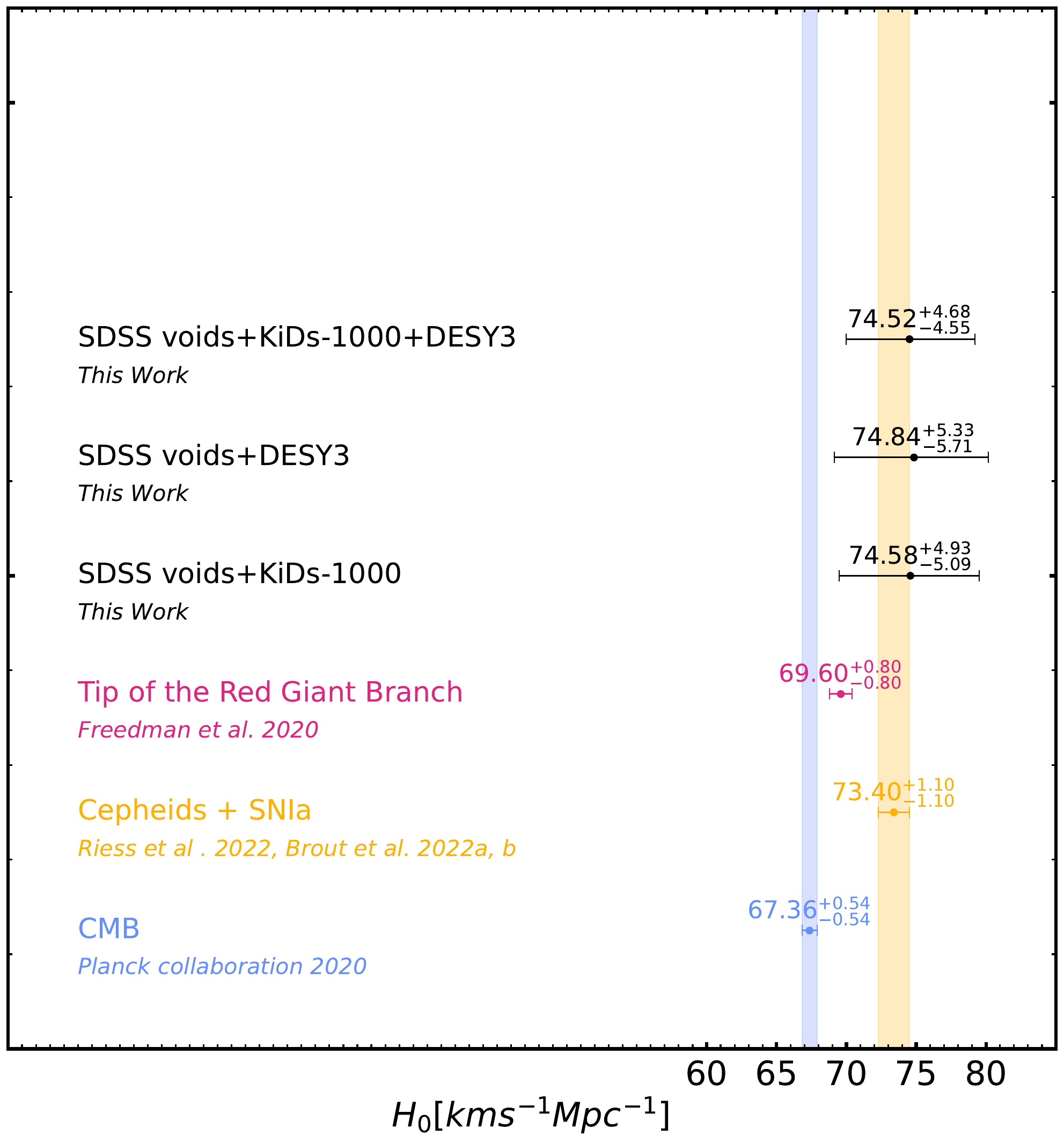}
    \caption{Comparison between recent constraints on the parameters $\sigma_{8}$ (left panel) and H$_{0}$ (right panel) from different cosmological probes. The error bars represent 68$\%$ confidence intervals. The black error bars represent the values constrained in this work. The references of the rest of the works, from top to bottom in left panel are:  \cite{2023OJAp....6E..36D}, \cite{refId0}, and \cite{planck2018}, and for the right panel:  \cite{Freedman_2020}, \cite{2022ApJ...934L...7R}, \cite{2022ApJ...938..110B, 2022ApJ...938..111B}, and \cite{planck2018}.} 
    \label{comparison}
\end{figure*}

\begin{figure}[tbh!]
    \centering
    \includegraphics[width=.5\textwidth]{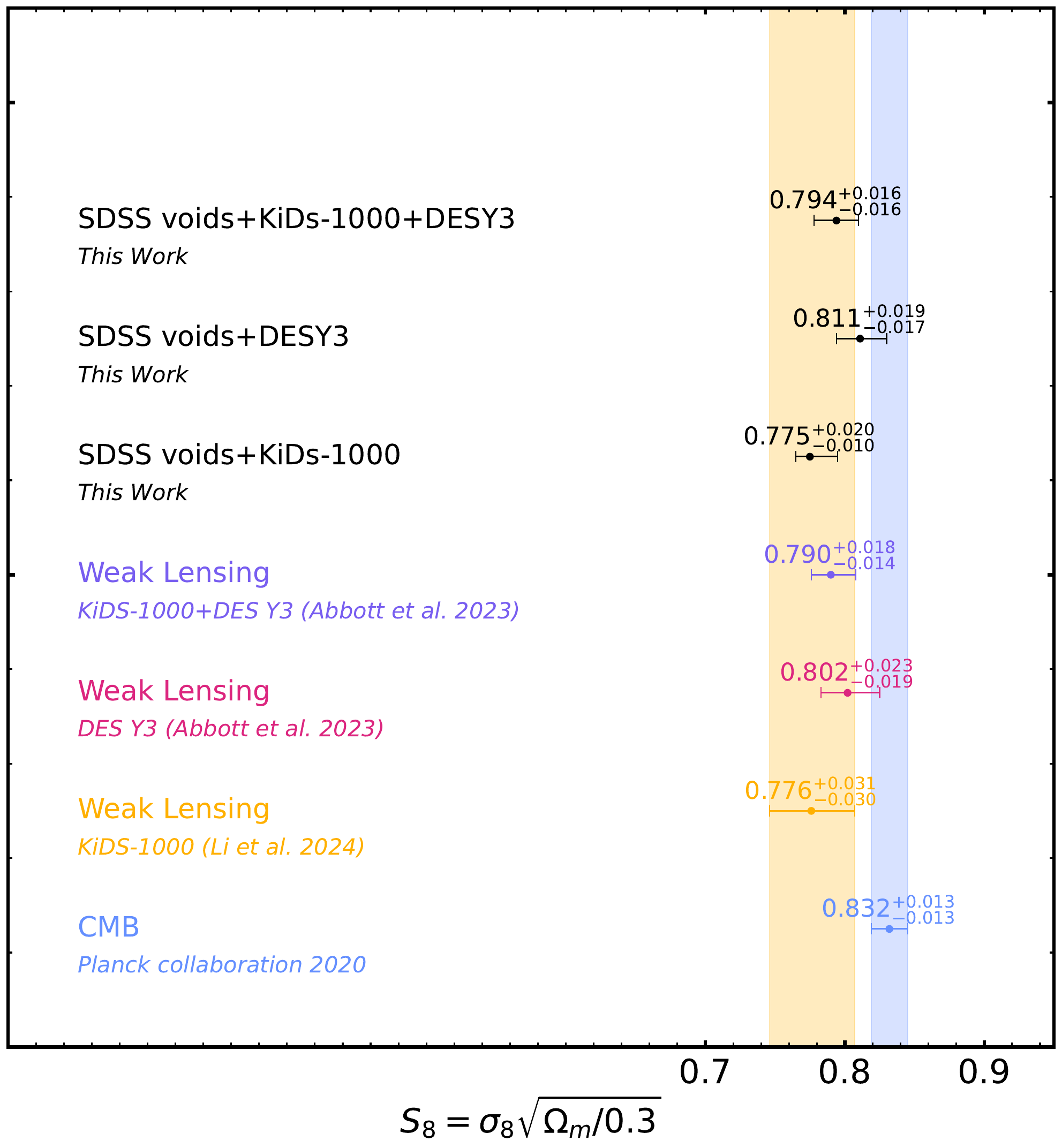}
    \caption{Comparison between recent constraints on S$_{8}$ from different cosmological probes. The error bars represent 68$\%$ confidence intervals. The black error bars represent the values constrained in this work. The references of the rest of the works, from top to bottom are: \cite{2023OJAp....6E..36D}, \cite{Li_2023}, and \cite{planck2018}.} 
    \label{comparisonS8}
\end{figure}

We now compare our results with those from CMB, weak lensing, Cepheid-based measurements, and tip of the red giant branch experiments. Figures \ref{comparison} and \ref{comparisonS8} display the constraints on $\sigma_{8}$, H$_{0}$, and S$_{8}$ obtained in this work alongside those from the aforementioned studies. As discussed in the introduction, there is a significant discrepancy between the constraints on S$_{8}$ (and $\sigma_{8}$) and H$_{0}$ derived from high-redshift and low-redshift cosmological probes. This discrepancy is evident in Figures \ref{comparison} and \ref{comparisonS8}, where it is clear that the constraints from CMB and weak lensing are not consistent with those from Cepheids and SNIa measurements.

In the $\sigma_{8}$ panel of Figure \ref{comparison}, we observe that combining our results with KiDS-1000 or KiDS-1000+DESY3 yields $\sigma_{8}$ values that are compatible with Planck 2018 within 1$\sigma$. However, combining our results with DESY3 alone produces a combined $\sigma_{8}$ value that is incompatible with both Planck 2018 and DESY3 within 1$\sigma$. Additionally, the combined H$_{0}$ values from all three weak lensing studies are consistent within 1$\sigma$ with the H$_{0}$ measurements obtained from Cepheids+SNIa.

In Figure \ref{comparisonS8}, we can see a comparison of the constrained S$_{8}$ values obtained by combining SDSS voids with KiDS-1000 or KiDS-1000+DESY3 with CMB and weak lensing constraints. We can see that the constraints from SDSS combined with the three weak lensing studies considered in this work are consistent within 1$\sigma$. When we combine SDSS voids with DESY3, the resulting S$_{8}$ value is also compatible within 1$\sigma$ with Planck 2018, but the rest of the combinations are not.

To conclude this section, we estimate the potential reduction in errors, considering only void statistics, with future redshift surveys such as Euclid \citep{2020A&A...642A.191E} and the Dark Energy Spectroscopic Instrument (DESI) experiment \citep{2016arXiv161100036D, 2016arXiv161100037D}. Euclid aims to investigate the Universe's expansion history and the evolution of large-scale structures by measuring the shapes and redshifts of galaxies, covering 15,000 deg$^{2}$ of the sky and extending to redshifts of approximately $z = 2$. 

Similarly, DESI will perform the Bright Galaxy Survey (BGS), targeting over 10 million galaxies in the redshift range $0 < z < 0.6$, as well as a dark-time survey of 20 million luminous red galaxies (LRGs), emission-line galaxies (ELGs), and quasars \citep{Hahn_2023}. With a projected footprint of 14,000 deg$^{2}$ and a broader redshift range, DESI is expected to achieve a precision 1–2 orders of magnitude higher than that of existing surveys, such as SDSS and BOSS.

We can estimate the reduction in uncertainties for the main parameters of the theoretical framework ($\sigma_{8}$ and $\Gamma$) when using DESI Y1 by calculating the ratio between the volumes of the SDSS and DESI surveys to the power of a constant value which has been calculated for each parameter using the constraints from a different number of lightcones (see Appendix \ref{scaling_errors} for more details). For $\sigma_{8}$, the uncertainties would decrease as 

\begin{equation}
    \Delta \approx \left(\frac{V_{BGS,Y1}}{V_{SDSS}}\right)^{-1/7} \sim \left(\frac{1.2\times10^{9} h^{-3} {\rm Mpc}^{3}}{40\times10^{6} h^{-3} {\rm Mpc}^{3}}\right)^{-1/7} \sim 0.62
;\end{equation}

in other words, the uncertainties for these two parameters would be 1.63 times smaller than the uncertainties obtained with the sample of SDSS used in this work. 

For $\Gamma$ the decrease would be 

\begin{equation}
    \Delta \approx \left(\frac{V_{BGS,Y1}}{V_{SDSS}}\right)^{-1/2.5} \sim \left(\frac{1.2\times10^{9} h^{-3} {\rm Mpc}^{3}}{40\times10^{6} h^{-3} {\rm Mpc}^{3}}\right)^{-1/2.5} \sim 0.26
,\end{equation}
which implies that the uncertainties for this parameter would be 3.9 times smaller than the uncertainties obtained with the sample of SDSS used in this work.

Therefore, the uncertainties for $\sigma_{8}$ (and S$_{8}$) obtained with the Bright Galaxy Survey from DESI Y1 will be approximately a factor 1.7 lower than those obtained in this work with the SDSS sample used in this work, and a factor 3.9 for $\Gamma$.  These uncertainties will be even lower when the full DESI survey is complete.

\section{Summary and future developments}\label{discussion}
In this work, we have made use of the theoretical framework developed in \cite{art5} and re-calibrated the expression for the number density of voids greater than $r$ (see Eq. (\ref{numbervoids})). We  used the Uchuu halo simulation box with a number density of halos equal to $3\times10^{-3}h^{3}$Mpc$^{-3}$, which has a greater volume than the simulation used in the aforementioned work ($V=2000^{3} h^{-3}$Mpc$^{3}$). The most important results obtained in this work are as follows:

\begin{itemize}
    \item We have proven a compatibility within $1\sigma$ among the number density of voids greater than $r$ 
    of SDSS galaxies with $M_{r}<-20.5$ (where $M_{r}$ is the absolute magnitude in $r-$band), $z_{min}=0.02$, and $z_{max}=0.132$, as well as Uchuu-SDSS galaxies with the same characteristics.     \item We have demonstrated that our theoretical framework  can successfully 
    predict the abundance of large voids for the four halo simulation boxes with different values of $\sigma_{8}$ ($\sigma_{8}=\{0.8159, 0.8102, 0.75, 0.65\}$) and Uchuu-SDSS lightcones used in this work,
    \item We  used a Bayesian analysis for all halo simulation boxes and  calculated the contours for 68$\%$ ($1\sigma$) and 95$\%$ ($2\sigma$) credible intervals in $\sigma_{8}$-H$_{0}$ plane (fixing $\Omega_{\rm m}$ in each case to its real value; see Table \ref{parametros_3}). We proved that we can recover the real values of $\sigma_{8}$ and H$_{0}$ of each halo simulation box within $1\sigma$ in real space and  $2\sigma$ in redshift space, 
    \item We  checked that using MCMC sampler from Cobaya, we successfully recover the values of these parameters for Uchuu-SDSS lightcones. The recovered values are $\sigma_{8}=1.296^{+0.621}_{-0.592}$, $\Omega_{\rm m}=0.419^{+0.172}_{-0.152}$, H$_{0}=72.35^{+12.24}_{-11.85}$, $\Gamma=0.250^{+0.096}_{-0.088}$, and S$_{8}=1.61^{+1.05}_{-0.93}$. 
    \item We  calculated the contours in $\sigma_{8}-\Omega_{\rm m}$, $\sigma_{8}$-H$_{0}$, and H$_{0}-\Omega_{\rm m}$ planes for the same ranges as mentioned above for SDSS void statistics. The contours obtained for SDSS are much broader than the ones obtained for halo simulation boxes (and Uchuu-SDSS lightcones) because of the huge difference of volume (the SDSS volume is approximately 25 times smaller than the three small simulation boxes and 200 times smaller than Uchuu box volume), which means that there are far fewer voids in SDSS than in the boxes. The constrained values obtained from SDSS are $\sigma_{8}=1.520^{+0.416}_{-0.441}$, $\Omega_{\rm m}=0.459^{+0.184}_{-0.184}$, H$_{0}=71.63\pm^{+12.60}_{-11.77}$, $\Gamma=0.270^{+0.094}_{-0.100}$, and S$_{8}$=1.87$^{+0.59}_{-0.76}$. It is important to remark that these constrained values have been obtained supposing that the ratio of $\Omega_{b}/\Omega_{\rm m}$ is constant and given by Planck 2018.
    \item Next, we  combined our SDSS voids constraints with KiDS-1000, DESY3, and KiDS-1000+DESY3 \citep{2023OJAp....6E..36D}. The results obtained when combining SDSS voids with KiDS-1000+DESY3 are:  $\sigma_{8}=0.858^{+0.040}_{-0.040}$, $\Omega_{\rm m}=0.257\pm^{+0.023}_{-0.020}$, H$_{0}=74.17^{+4.66}_{-4.66}$, and S$_{8}$=0.794$^{+0.016}_{-0.016}$. We  see that when we combine our results with weak lensing works, we significantly improve  the uncertainties of all the parameters (with the exception of H$_{0}$) with respect to the uncertainties of only KiDs-1000+DESY3. This is because the marginalised posteriors in the  $\sigma_{8}-\Omega_{\rm m}$ plane obtained in this work and those from KiDS-1000 are almost orthogonal.
    \item We have seen that any value of H$_{0}$ or $\Omega_{\rm m}$ obtained when combining SDSS voids with the three weak lensing works are compatible with Planck 2018 within 1$\sigma$ (for $\Omega_{\rm m}$, they are within approximately 2$\sigma$). We have also seen that S$_{8}$ value from SDSS voids + DESY3 is compatible with Planck within 1$\sigma$, but SDSS voids + KiDS-1000 and SDSS voids + KiDS-1000 + DESY3 are not (the former is compatible within 2.7$\sigma$ and the latter within 2$\sigma$).
    \item Finally, we  compared our results with those obtained in other works where voids have also been used to constrain the same cosmological parameters constrained in the current work, along with parameters constrained via CMB, galaxy clustering, weak lensing, and Type Ia supernova measures. We have checked that when our sample is combined with those of weak lensing studies, we obtained slightly smaller uncertainties than in other works focussed solely on voids. However, when we do not combine our results with any work, our uncertainties are very large because the volume of the samples of Uchuu-SDSS and SDSS used in this work is relatively small, in comparison to the volumes of redshift surveys used in other works of voids. For example, in  \cite{Contarini_2024}, the BOSS redshift survey was used. If we used this redshift survey, we would have predicted that our uncertainties (without combining with any work) for $\sigma_{8}$ and S$_{8}$ would be approximately a 30$\%$ larger than the ones given in \cite{Contarini_2024} (where the authors combined void statistics with void shapes).
    \item Therefore, as we have demonstrated in Section \ref{comparison_sec}, the most important limitation we face when constraining the cosmological parameters considered in this study is the small volume of the redshift survey sample we use, which is approximately $\sim 40\times10^{6} h^{-3}$Mpc$^{3}$. When we use a survey such as DESI Y1, we would have estimated uncertainties for $\Gamma$ ($\sigma_{8}$) that are approximately 4 (1.5) times smaller than the uncertainties obtained with the sample of SDSS used in this work.  If we  combined our study with other, additional works, then the decrease in the errors would be greater, as a result of increasing the volume of the sample of galaxies used.  
\end{itemize}

\section*{Data availability}
The Uchuu halo and galaxy boxes and the 4 box catalogs at redshift z=0.092, as well as the 32 Uchuu-SDSS galaxy lightcones, the SDSS catalogue, and the void catalogues from all the previous galaxy and halo catalogues used in this work are available at: \url{http://www.skiesanduniverses.org/Simulations/Uchuu/}. This link includes information on how to read the data and column description. For a list and brief description of the available halo, Uchuu-SDSS and SDSS void catalogues columns, see Appendix \ref{columns}.

\begin{acknowledgements}
      E. Fernández-García acknowledges financial support from the Severo Ochoa grant CEX2021-001131-S funded by MCIN/AEI/ 10.13039/501100011033.
      EFG, FP and AK thanks support from the Spanish MICINN PID2021-126086NB-I00 funding grant.
T.I has been supported by IAAR Research Support Program in Chiba University Japan, MEXT/JSPS KAKENHI (Grant Number JP19KK0344, JP21H01122, and JP23H04002),
MEXT as ``Program for Promoting Researches on the Supercomputer Fugaku'' (JPMXP1020200109 and JPMXP1020230406), and JICFuS.
    We thank Instituto de Astrofisica de Andalucia (IAA-CSIC), Centro de Supercomputacion de Galicia (CESGA) and the Spanish academic and research network (RedIRIS) in Spain for hosting Uchuu DR1, DR2 and DR3 in the Skies $\&$ Universes site for cosmological simulations. The Uchuu simulations were carried out on Aterui II supercomputer at Center for Computational Astrophysics, CfCA, of National Astronomical Observatory of Japan, and the K computer at the RIKEN Advanced Institute for Computational Science. The Uchuu Data Releases efforts have made use of the skun$@$IAA$\_$RedIRIS and skun6$@$IAA computer facilities managed by the IAA-CSIC in Spain (MICINN EU-Feder grant EQC2018-004366-P).
    The other cosmological simulations were carried out on the supercomputer Fugaku provided by the RIKEN Center for Computational Science (Project ID: hp220173, hp230173, and hp230204).
    We also thank Joe Zuntz and Anna Porredon for their feedback on our cosmological results. Finally, we are thankful to Peter Taylor for instructing us on the the use of \texttt{CombineHarvesterFlow}, for verifying our results and for giving us feedback on the paper. 
\end{acknowledgements}


%
   \bibliographystyle{aa} 
   \bibliography{aanda} 
%

\begin{appendix}

\section{Void finder code description}\label{voidfinderalgorithm}

\begin{figure}[tbh!]
    \centering
    \includegraphics[width=.5\textwidth]{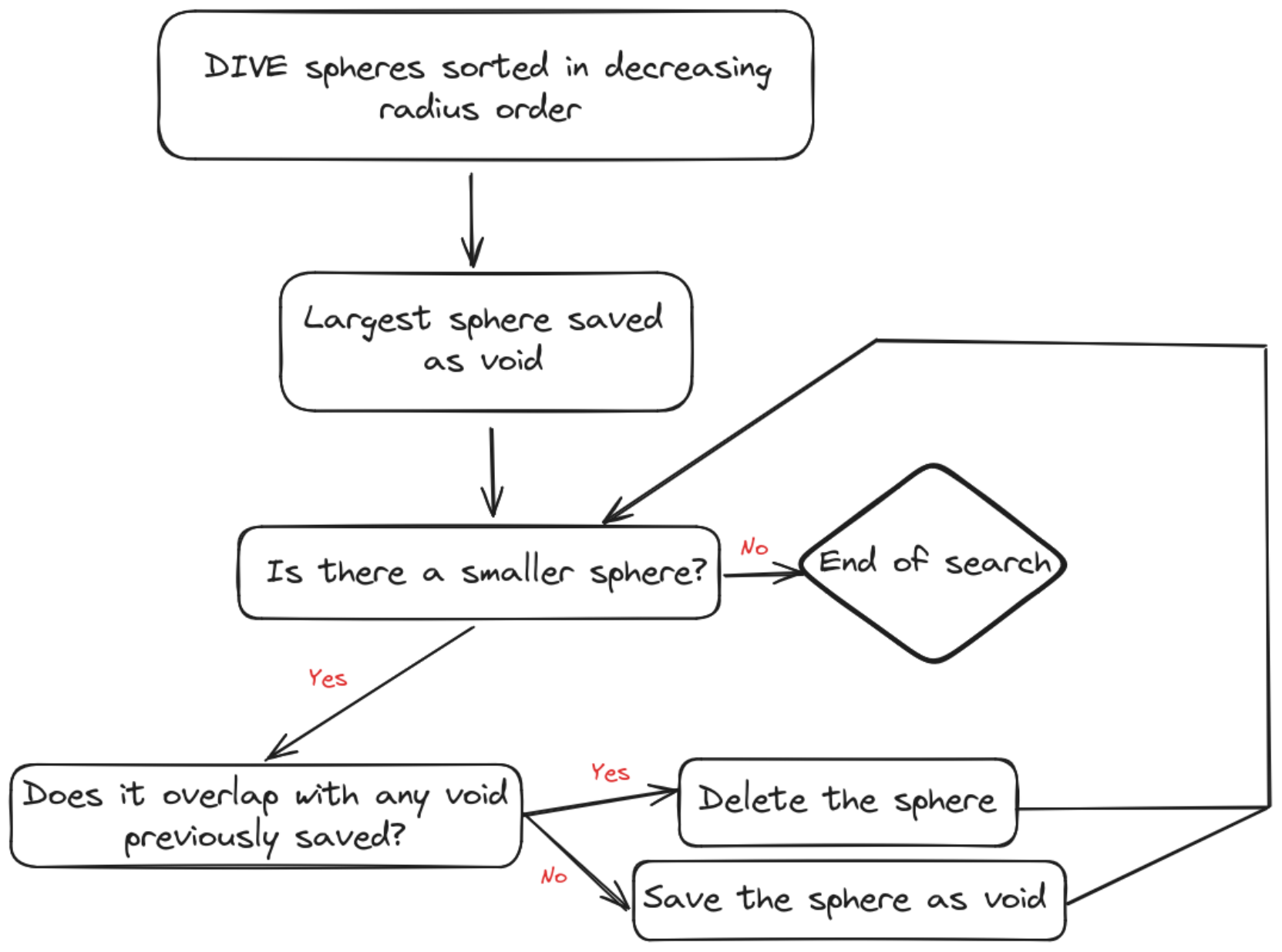}
    \caption{Void identification process.}
    \label{voids_code_diagram}
\end{figure}

In this appendix, we outline the custom \texttt{C++} code developed to identify maximal non-overlapping spheres (voids) from the candidates provided by \texttt{DIVE}. The code utilizes \texttt{OpenMP} for parallelization. A summary of the implementation and its workflow is illustrated in Figure \ref{voids_code_diagram}.

The code begins by including necessary headers and defining a structure for spheres with center coordinates $(X, Y, Z)$ and radius $R$. It also includes a utility function to check overlaps: two spheres, $s_1$ and $s_2$, overlap if the distance between their centers satisfies:
\begin{equation} (X_1-X_2)^2 + (Y_1-Y_2)^2 + (Z_1-Z_2)^2 < (R_1+R_2)^2 \label{eq:check_overlap} \end{equation}
The \texttt{main} function performs the following steps:
\begin{enumerate}
    \item Reads the input file containing DIVE candidates.
    \item Sorts candidates by descending radius, $R$.
    \item Initializes the void list with the largest candidate.
    \item terates through the remaining candidates, checking overlaps using Eq. (\ref{eq:check_overlap}):
    \begin{itemize}
        \item If no overlap is detected, the candidate is retained as a void.
        \item If overlap is detected, the candidate is discarded.
    \end{itemize}
    \item Outputs the identified voids to a file.
\end{enumerate}

The resulting output provides the centers and radii of all voids found in the halo or galaxy sample.

\section{Void statistics theoretical framework in detail}\label{computation}
In this appendix we explicitly write all necessary quantities in order to calculate $P_{0}(r)$ (VPF) and $\bar{n}_{v}(r)$. This theoretical framework can also be consulted in \cite{art5}.
The first step to calculate all the terms that appear in Eq.  (\ref{numbervoids}) is to develop Eq.  (\ref{Pn}). In \cite{patiri}, it is shown that for dark matter halos, $u$ can be written as
\begin{equation}
    u=\left[\bar{n}V(1+\delta)\right]\left[1+\delta_{ns} \right]
,\end{equation}
where $\bar{n}$ denotes the mean number density of those halos in the sample (usually halos larger than some given mass), $V$ is the volume of the sphere and $\delta$ the actual enclosed density contrast within the sphere. The first factor on the right-hand side of the equation is the integral of the probability density within the sphere for halos tracing the mass (i.e. no bias, which is true in the very low mass limit). In general, halos are biased tracers of the underlying mass distribution, due to the initial clustering of the protohalos before they move along with mass (i.e. statistical clustering). The second term of the equation accounts for this biasing. In \cite{patiri} an approximation for this bias is obtained as a function of the linear enclosed density contrast within the sphere ($\delta_{l}$):

\begin{equation}
    1+\delta_{ns}(\delta_{l}) = A(m)e^{-b(m)\delta_{l}^{2}} \hspace{10pt} \forall \hspace{10pt} \delta_{l} \leq -1
    \label{deltans}
,\end{equation}

where $A(m)$, $b(m)$ are coefficients mainly depending on the halo mass. In \cite{patiri} is demonstrated that, using Zeldovich approximation, $P_{n}(r)$ can be rewritten as

\begin{equation}
    P_{n}(r) = \int_{-\infty}^{1.6}P(\delta_{l}, r) \frac{\left[u(\delta_{l}) \right]^{n}}{n!}e^{[-u(\delta_{l})]}d\delta_{l}
,\end{equation}
where $u(\delta_{l})$ is now a function of $\delta_{l}$ through the dependence of the actual density contrast, $\delta$, on its linear counterpart, $\delta_{l}$:

\begin{equation}
    u(\delta_{l}) = (\bar{n}V[1+\delta(\delta_{l},r)])[1+\delta_{ns}(\delta_{l})]
,\end{equation}
where $\delta(\delta_{l},r)$ is basically the relationship between the actual and the linear density contrast within a sphere as given by the standard spherical collapse model, except for a small correcting term depending on $r$. 

Doing some manipulation, $P_{n}(r)$ can be rewritten as

\begin{equation}
        P_{n}(r) = \frac{1}{n!}\int_{-7}^{0}P(\delta_{l}, r)\left[u(\delta_{l}) \right]^{n}e^{[-u(\delta_{l})]}d\delta_{l}
\label{appendix_eqPn}
,\end{equation}
where $u(\delta_{l})$ is now

\begin{equation}
    u(\delta_{l}) = \bar{n}V[1+DELF(\delta_{l},r)]Ae^{-b\delta_{l}^{2}}
    \label{a2}
.\end{equation}
DELF is a function of $\delta_{l}$ and $r$ that gives the mean actual density contrast within a sphere with radius $r$ with enclosed linear density contrast $\delta_{l}$:

\begin{equation}
\begin{aligned}
     DELF(\delta_{l}, r) &= \\ \frac{1+DELT(\delta_{l})}{|1-(4/21)[1+DELT(\delta_{l})]^{2/3}[\sigma(r[1+DELT(\delta_{l})]^{1/3})]^{2}|}
\end{aligned}
\label{delf}
,\end{equation}

where $DELT(\delta_{l})$ denotes the relationship between the actual and linear enclosed density contrasts in the spherical collapse model \citep{patiri}:

\begin{equation}
    1+DELT(\delta_{l}) \approx (1-0.607\delta_{l})^{-1.66}
,\end{equation}

where $\sigma(Q)$ is the rms of the linear density contrast on a sphere with Lagrangian radius $Q$. In this equation, $\sigma(Q)$ is evaluated at $Q=r[1+DELT(\delta_{l})]^{1/3}$. $A$, $b$ in Eq. (\ref{a2}) are also functions of $\delta_{l}$  \cite[see][for more details]{2008MNRAS.386.2181R} given by:

\begin{equation}
\begin{aligned}
    A &= A(m, Q = r[1+DELT(\delta_{l})]^{1/3})\times \\ &\times\left[\frac{D(z)\sigma_{8}}{0.9} \right]^{0.88}\left(\frac{\Gamma}{0.21} \right)^{0.174}
    \label{Asigma}
\end{aligned}
,\end{equation}

\begin{equation}
\begin{aligned}
    b &= b(m, Q = r[1+DELT(\delta_{l})]^{1/3})\times \\ &\times\left[\frac{D(z)\sigma_{8}}{0.9} \right]^{-2.55}\left(\frac{\Gamma}{0.21} \right)^{-0.69}
    \label{Bsigma}
\end{aligned}
,\end{equation}
where $A(m,Q)$, $b(m,Q)$ are functions of the mass of the objects and the Lagragian radius of the regions being considered:

\begin{equation}
\begin{aligned}
    A(m,Q) &= \left[1.577-0.298\left(\frac{Q}{8}\right) \right] -\\ &- \left[0.0557+0.0447\left(\frac{Q}{8}\right) \right]\ln{m} -\\ &- \left[0.00565+0.0018\left(\frac{Q}{8}\right) \right][\ln{m}]^{2}
    \label{Aeq}
\end{aligned}
,\end{equation}

\begin{equation}
\begin{aligned}
    b(m,Q) &= \left[0.0025-0.00146\left(\frac{Q}{8}\right) \right] +\\ &+ \left[0.121-0.0156\left(\frac{Q}{8}\right) \right]m^{0.335+0.019Q/8}
    \label{beq}
\end{aligned}
,\end{equation}

\begin{equation}
    m = \frac{1.05M}{3.4866\times 10^{11} h^{-1}M_{\odot}}\left(\frac{0.3}{\Omega_{\rm m}} \right)
    \label{masnorm}
,\end{equation}
where $D(z)$ is the linear growth factor normalised to be 1 at present and $M$ is the minimum mass of halos when the matter density field is traced by the halo positions (or central halos). However, the matter density field of Uchuu is traced by the positions of its sub-structures (or subhalos), so the mass to be put in the theoretical framework is not $m$, but $m_{g}$ (see Appendix \ref{msubg_ex} for a detailed explanation of how one can calculate $m_{g}$).

We already have all terms in order to calculate $u(\delta_{l})$ from Eq.  (\ref{appendix_eqPn}), so all we need now is to calculate $P(\delta_{l}, r)$, which is the probability function of the linear density contrast within an Eulerian space, and can be calculated as \citep{Betancort-Rijo_2002}:

\begin{equation}
\begin{aligned}
    P(\delta_{l},r) &= \frac{exp[(-1/2)\delta_{l}^{2}/(\sigma(r[1+DELF(\delta_{l},r)]^{1/3}))^{2}]}{\sqrt{2\pi}}\times\\ &\times[1+DELF(\delta_{l},r)]^{-[1-(\alpha/2)]}\times \\ &\times\frac{d}{d\delta_{l}}\left(\frac{\delta_{l}}{\sigma(r[1+DELF(\delta_{l},r)]^{1/3})} \right)
    \end{aligned}
,\end{equation}
where
\begin{equation}
    \alpha(\delta_{l},r)=0.54+0.173ln\left(\frac{r[1+DELF(\delta_{l},r)]^{1/3})}{10}\right)
,\end{equation}
and
\begin{equation}
    \sigma(Q) = \sigma(Q,\Gamma) \approx \sigma_{8}A(\Gamma)Q^{-B(\Gamma)-C(\Gamma)Q}
    \label{sigmaeq}
,\end{equation}
\begin{equation}
    A(\Gamma)=2.01+3.9\Gamma
,\end{equation}
\begin{equation}
    B(\Gamma)=0.2206+0.361\Gamma^{1.5}
    \label{Beq}
,\end{equation}
\begin{equation}
    C(\Gamma)=0.182+0.0411\ln{\Gamma}
    \label{Ceq}
.\end{equation}
This fit is valid for $Q\geq3h^{-1}$Mpc and $0.1\geq\Gamma\geq0.5$. 

It is important to remark that Eq.  (\ref{sigmaeq}) is only valid for $z=0$. If we want to calculate this function in a different redshfit, then we have to make the following change: 

\begin{equation}
    \sigma_{8} \rightarrow \sigma_{8}\frac{D(z)}{D(z=0)}
    \label{sigma8redshift}
,\end{equation}
where $D(z)$ is the linear growth factor of density fluctuations in the model under consideration.

Finally, it's important to write the equation for the root mean square (rms) of the error of the estimation of the VPF. The expression we use is the following:
\begin{equation}
     rms(P_{0}(r))^{2} = \frac{9.2\left[1-\omega\bar{n}_{v}(r)\right]P_{0}(r)^{2}}{N(r)}+\frac{P_{0}(r)}{N_{spheres}}
     \label{rmsPo}
,\end{equation}
where $N_{spheres}$ is the number of random spheres used in order to estimate $P_{0}(r)$ for simulations, $N(r)$ is the number of voids larger than $r$. This equation is a modification of the expression given in \cite{1992PhRvA..45.3447B} and \cite{10.1111/j.1365-2966.2006.10305.x}. We have added a second term, $P_{0}(r)/N_{spheres}$ to take into account the error due to the finite number of trial random spheres used to calculate the VPF. The first term of Eq.  (\ref{rmsPo}) takes into account the finite volume of the sample.

Then, $\omega$ in Eq.  (\ref{rmsPo}) is given by

\begin{align}
    \omega = \frac{32\pi}{3}\bar{R}^{3}&\left[1+2.73\bar{n}_{v}(r)\frac{32\pi}{3}\bar{R}^{3}\times \right. \nonumber \\
    &\times\left.\left(1-\frac{3}{4}\frac{\bar{n}_{v}(r)^{-1/3}}{2\bar{R}}+\frac{1}{8}\left(\frac{\bar{n}_{v}(r)^{1/3}}{2\bar{R}}\right)^{2}\right)\right]^{-1},
\end{align}
where $\bar{R}$ is the mean radius of all voids larger than the minimum voids considered.

\section{Calculation of m$_{g}$}\label{mass_detailed}
As shown in Appendix \ref{computation}, $P_{n}(r)$ (in concrete, $u$), depends on $m$ ($M$ normalised according to Eq.  (\ref{masnorm})), which is the minimum mass of halos when the matter density field is traced by the halo positions. However, we want to use this theoretical framework to infer cosmological parameters from Uchuu, which matter density field is traced by its sub-structures (subhalos), and from SDSS galaxies, which are also substructures of halos. In this case, the mass to be used in the theoretical framework is not $m$ but $m_{g}$ (i.e. the minimum mass of subhalos --including halos).

These two masses, $M$ and $M_{g}$, can be calculated directly from the simulations used. For example, $M$ can be calculated imposing $\bar{n}(>M)=\bar{n}_{\rm sample}$, where $\bar{n}_{\rm sample}$ is the number density of the sample considered. In our case, $\bar{n}_{\rm sample}$ = 3$\times10^{-3}$ $h^{3}$Mpc$^{-3}$. Note that in this case $\bar{n}(>M)$ is the cumulative halo mass function. $M_{g}$ can be calculated the same way, but taking also into account the subhalos in $\bar{n}(>M_{g})$. 

Therefore, in a simulation such as Uchuu, with halos and subhalos, we can calculate both $M$ and $M_{g}$ and establish a relationship between both. This was already done in \cite{art5} (see Eq.  (16) from that article). However, in this work we have recalibrated this expression and found that $m_{g}$ can be approximately related to $m$ through the following equation:

\begin{equation}
    m_{g} = 1.058^{\sigma(m)}m
    \label{msubg_eq}
,\end{equation}
where $\sigma(m)$ is the rms linear density fluctuation on scale $m$. It can be shown that, in the interval of masses and cosmological parameters considered in this work, $\sigma(m)$ depends on $m$, $\Gamma$ and $\sigma_{8}$ the following way:

\begin{equation}
    \sigma(m)=2.035\left(\frac{m}{4.039} \right)^{-\frac{0.475}{3}}\left(\frac{\Gamma}{0.1754} \right)^{0.27}\left(\frac{\sigma_{8}}{0.8159} \right)
,\end{equation}

Additionally, an expression for $m_{g}$ depending on the $\bar{n}_{sample}$, $\sigma_{8}$ and $\Gamma$ can be found:

\begin{equation}
    m_{g} = 4.74\left(\frac{\sigma_{8}}{0.8159} \right)^{0.402}\left(\frac{\Gamma}{0.1754} \right)^{0.109}\left(\frac{F(\bar{n}_{sample
    })}{4.047} \right)
    \label{msubg_ex}
,\end{equation}
where

\begin{equation}
    F(\bar{n}) = -\frac{1.267\times10^{-8}}{\bar{n}^{2}}+\frac{1.289\times10^{-2}}{\bar{n}}-0.248
.\end{equation}

This last function, $F(\bar{n})$  defines the dependence of the mass $m$ with the number density of the sample, $\bar{n}$. It is important to remark that $\bar{n}_{sample}$ is the number density of galaxies corrected from completeness in the case we use redshift surveys or lightcones. 

From Eq. (\ref{msubg_ex}), we can see that we obtain $m_{g}=4.74$ for Uchuu, which is the mass obtained using directly the simulations (for $n_{g}=3\times10^{-3}$h$^{3}$Mpc$^{-3}$, we take the the $N=n_{g}V$ most massive halos, where $V$ is the Uchuu volume, and fix $m_{g}$ to the minimum mass of this $N$ halos) using the normalisation given by Eq.  (\ref{masnorm}). This equation is a good approximation to $m_{g}$ as a function of cosmological parameters and $\bar{n}_{sample}$, while the abundance matching provides it exact value. However, the dependence on $\sigma_{8}$ and $\Gamma$ is quite small and it is swamped by small error in expressions (\ref{Asigma}-\ref{beq}) (which have been fitted to the results of a complex procedure) that have a much stronger dependence on $\sigma_{8}$ and $\Gamma$. Thus, we have found that the best agreement between the theoretical framework and the simulations are obtained holding $m_{g}$ fixed for given values of $\bar{n}_{sample}$.

The value of $m_{g}$ for the four halo boxes has been calculated using the chi square test with the VPF defined in real space\footnote{the value of $m_{g}$ must be the same in real and redshift spaces} for $r>14h^{-1}$Mpc, $r<21h^{-1}$Mpc and $\Delta r=1h^{-1}$Mpc (i.e 8 bins). We have minimised the sum of $\chi^{2}=\chi^{2}_{\rm Uchuu}+\chi^{2}_{\rm P18}+\chi^{2}_{\rm P18LowS8}+\chi^{2}_{\rm P18VeryLowS8}$, and have obtained an optimal value of $m_{g}=$4.90 with $\chi^{2}=41.67$ and  $\chi^{2}_{\nu}=\chi^{2}/\nu$=1.49 (where $\nu$ is the degree of freedom, $\nu=4\times7$ in our case). Therefore, $m_{g}=$4.90 is the value we will use in this work.

\section{Validation of the theoretical framework in real and redshift space}\label{validation_model}
As already mentioned, the theoretical framework predicts the void probability function (VPF) in real and redshift space from first principles and establish a relationship between the VPF and the number density of voids larger than $r$. This relationship between the VPF and the number density of voids larger than $r$ does not arise from first principles but is an assumption that, although not exact, is quite accurate (see Appendix \ref{calibration_model} for more details).

In Section \ref{sec2} and Appendix \ref{computation}, we give all the expressions needed to compute the VPF and $m_{g}$, although in Appendix \ref{mass_detailed} we calculate an universal mass for the 4 simulation boxes used in this work as the dependence of $m_{g}$ with $\sigma_{8}$ is small.

\begin{figure}[h!]
    \centering
    \includegraphics[width=.5\textwidth]{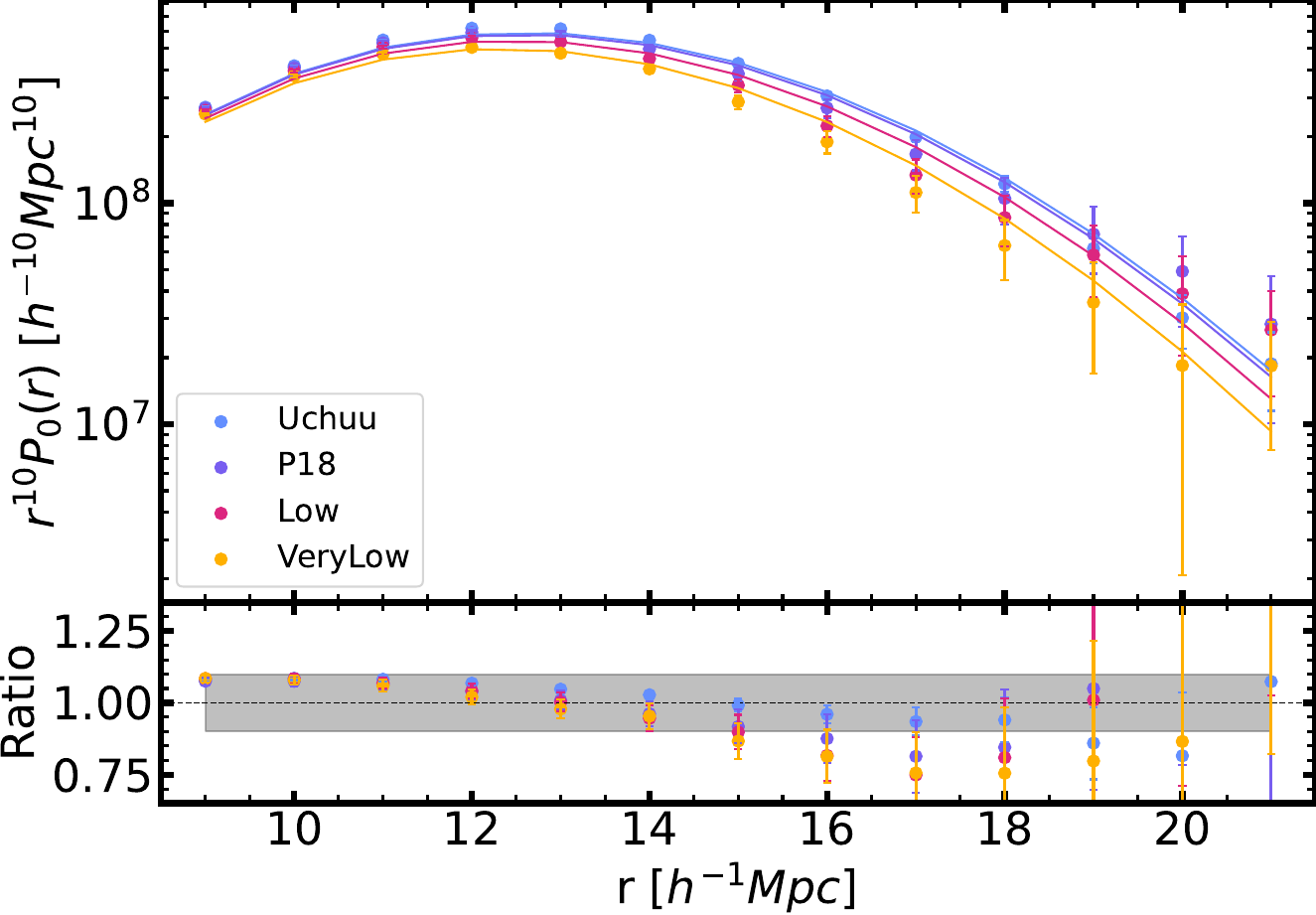}    
    \caption{ VPF in real space is shown for the theoretical framework (lines) and simulations (dots) given in the top panel, while the ratio between simulations and theoretical framework is shown in bottom panels for the Uchuu, P18, Low, and VeryLow box catalogues with number density $\bar{n}=3\times10^{-3} h^{3}$Mpc$^{-3}$. Shaded region in bottom panel indicates delimits the region between 0.9 and 1.10 for the ratio.}
    \label{VPF_halos}
\end{figure}

\begin{figure}[bth!]
    \centering
    \includegraphics[width=.5\textwidth]{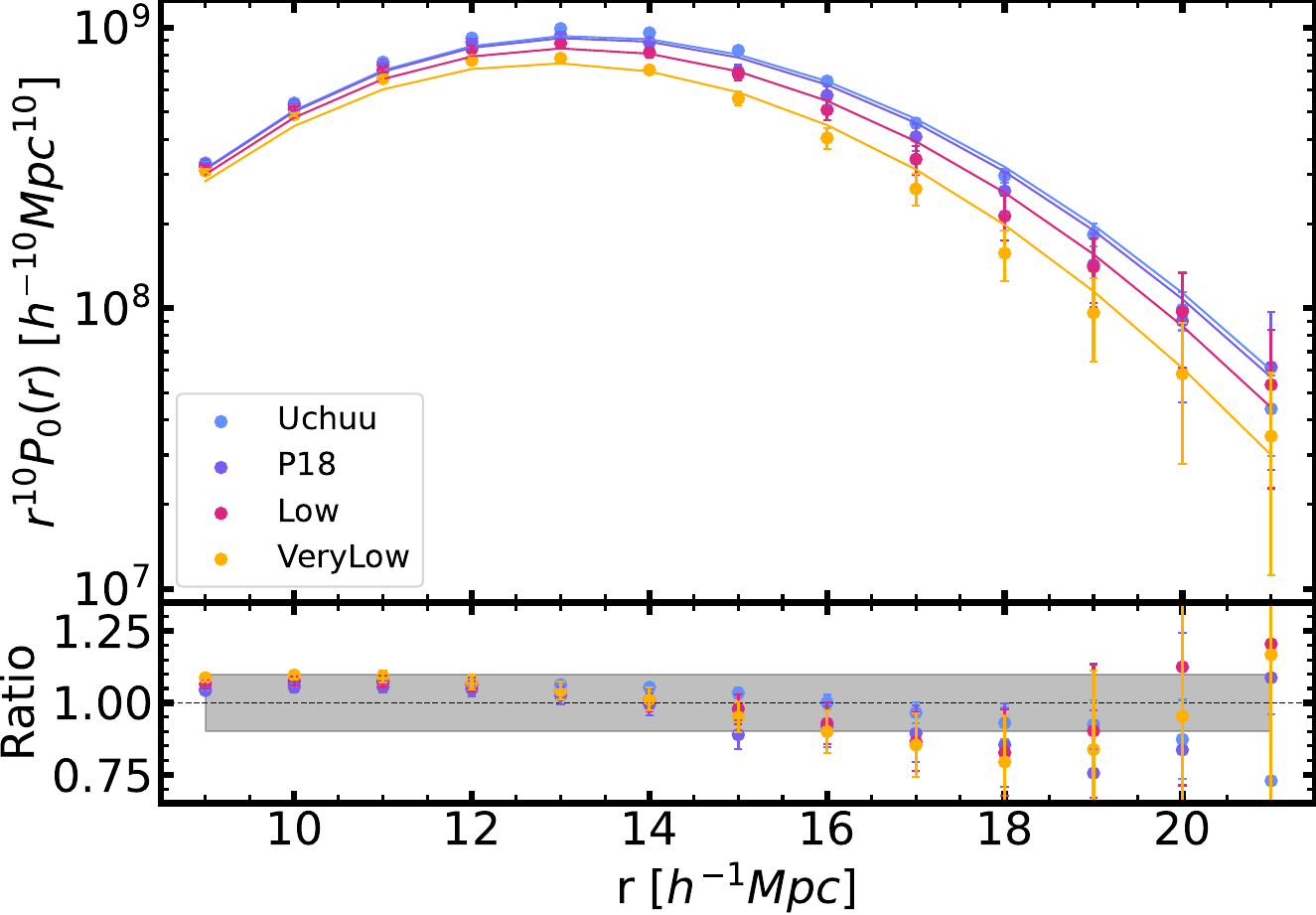}
    \caption{ VPF in redshift space is shown for the theoretical framework (lines) and simulations (dots) given in the top panel, while the ratio between simulations and theoretical framework is shown in bottom panels for the Uchuu, P18, Low and VeryLow box catalogues with number density $\bar{n}=3\times10^{-3} h^{3}$Mpc$^{-3}$. Shaded region in bottom panel indicates delimits the region between 0.9 and 1.10 for the ratio.}
    \label{VPF_halos_redshift}
\end{figure}

Therefore, we are in the position of checking if the VPF can reproduce the VPF from the 4 simulation boxes in real and redshift space. This can be seen in Figure \ref{VPF_halos} (real space) and \ref{VPF_halos_redshift} (redshift space). We can see that the values given by all the simulations are compatible within 1$\sigma$ with the values predicted by the theoretical framework, both in real and redshift space. Therefore, we can conclude that the theoretical framework, which predicts the VPF based on first principles, is valid.

\section{Calibration of the relationship between the VPF and the number density of voids larger than $r$}\label{calibration_model}
The theoretical framework presented in this work is based in the model presented in \cite{art5}. This model was calibrated with Millennium Run simulations \citep{2005Natur.435..629S}, which follows the evolution of 1010 dark matter particles in a periodic box of 500 h$^{-1}$ Mpc on a side with a mass resolution per particle of 8.6$\times$10$^{8}$ h$^{-1}$  M$_{\odot}$. This volume is extremely small in comparison with Uchuu's volume (Uchuu's volume is 64 times larger than Millennium Run's), so the void statistics calculated with this box weren't as accurate as the ones obtained with Uchuu, which translates in an imprecision in the calibration of the theoretical framework. 

The parameters of the theoretical framework that were calibrated with simulations were $\mu$, $\alpha$ and $\beta$, defined by

\begin{equation}
    \bar{n}_{v}(r) = \frac{\mu\mathcal{K}(r)}{V}e^{-\alpha\mathcal{K}(r)[1-\beta K(r)]}
,\end{equation}

along with the mass, $m_{g}$, to be used in all the halo and galaxy boxes,  as well as lightcones and SDSS. The calculation of $m_{g}$ is explained in Appendix \ref{msubg_ex}. 

In order to find the values of the coefficients $\mu$, $\alpha$, and $\beta$, we have used the chi square test with the number density of voids larger than $r$ predicted by the theoretical framework used in this work, letting these coefficients be free parameters, and the values given by the simulations. We have minimised sum of the chi square values of each halo box: $\chi^{2}=\chi^{2}_{\rm Uchuu}+\chi^{2}_{\rm P18}+\chi^{2}_{\rm P18LowS8}+\chi^{2}_{\rm P18VeryLowS8}$ and we have obtained the optimal value of $\mu=$0.603, $\alpha=$1.792 and $\beta=$ 0 with $\chi^{2}=26.99$ and  $\chi^{2}_{\nu}=\chi^{2}/\nu$=1.12 (where $\nu$ is the degree of freedom, $\nu=4\times6$ in our case). Therefore, these are the values of $\mu$, $\alpha$, and $\beta$ that we use in this work.

\section{Constraints of cosmological parameters using voids statistics in the distribution of halos and galaxies in mocks with theoretical framework}\label{appendixhaloes}
In this appendix, we show the marginalised posteriors obtained with the theoretical framework presented in Appendix \ref{computation} using the maximum likelihood test with Bayesian approach. We compare it with the results of the halo simulation boxes. We do this in real and redshift space. 

\subsection{Real Space}
\begin{figure}[b!]
    \centering
    \includegraphics[width=0.5\textwidth]{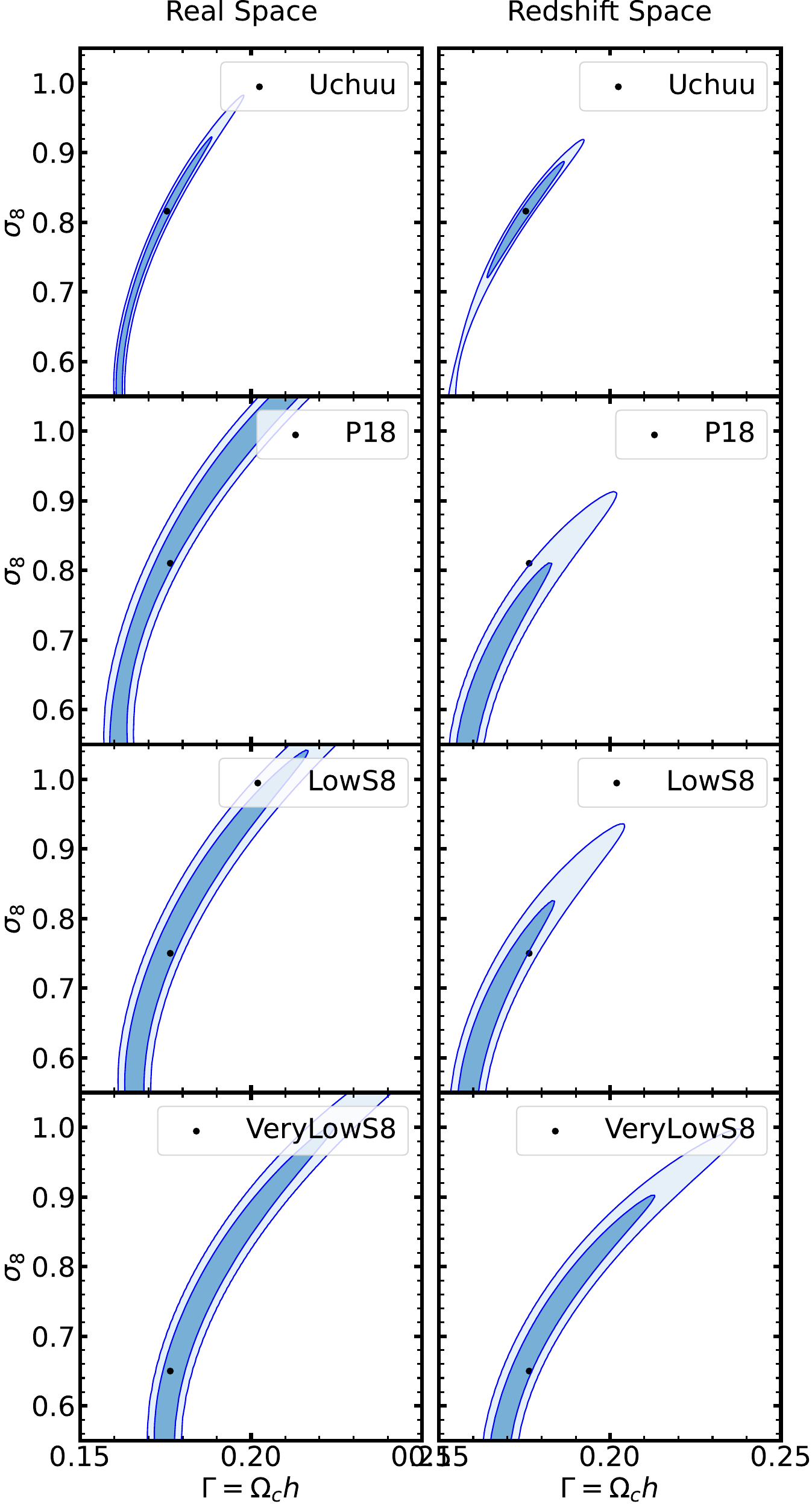}
    \caption{Constraints from $N_{v,i}$ identified in  Uchuu (first row), P18 (second row), Low (third row) and VeryLow (row) halo simulation boxes in real (first column) and redshift (second column) spaces using Maximum Likelihood test. The contours indicate the 68$\%$ (1$\sigma$) and 95$\%$ (2$\sigma$) credible intervals. The black dots are the real values of $\sigma_{8}$ and H$_{0}$ of each simulation.}
    \label{chi_cuadrado}
\end{figure}

We can see the marginalised posteriors obtained in the for halo boxes in real space in the first column of Figure \ref{chi_cuadrado}. In this figure, we can see that all fiducial values (black points) are inside the 1$\sigma$ contour. 
Therefore, we successfully recover the values of $\sigma_{8}$ and $\Gamma$ of the simulations with the theoretical framework, and we can take one more step to do this study in redshift space.

\subsection{Redshift space}\label{redshiftmocks}
The marginalised posteriors obtained with maximum likelihood test for the halo simulation boxes in redshift space can be seen in the second column of Figure \ref{chi_cuadrado}. In that figure, we can see that we obtain a small contour for Uchuu in redshift space, unlike in real space. Additionally, we recover Planck 2018 values of $\sigma_{8}$ and H$_{0}$ within $1\sigma$ for the P18 box catalogue, and within $2\sigma$ for Low and VeryLow. Therefore, in redshift space we have not successfully recovered the real values of the parameters in simulations within $1\sigma$ as we have in real space. This can be due to the transformation we have used in order to transform the theoretical framework from real to redshift space (see Eq. (\ref{rredshift})).  However, we can recover the real values within $2\sigma$ for the four halo simulation boxes.

\section{Scaling of errors with volume}\label{scaling_errors}
An important aspect to consider is the scaling of errors with the sample volume, specifically: how the errors decrease as the sample volume increases.

\begin{figure}[th!]
    \centering
    \includegraphics[width=.45\textwidth]{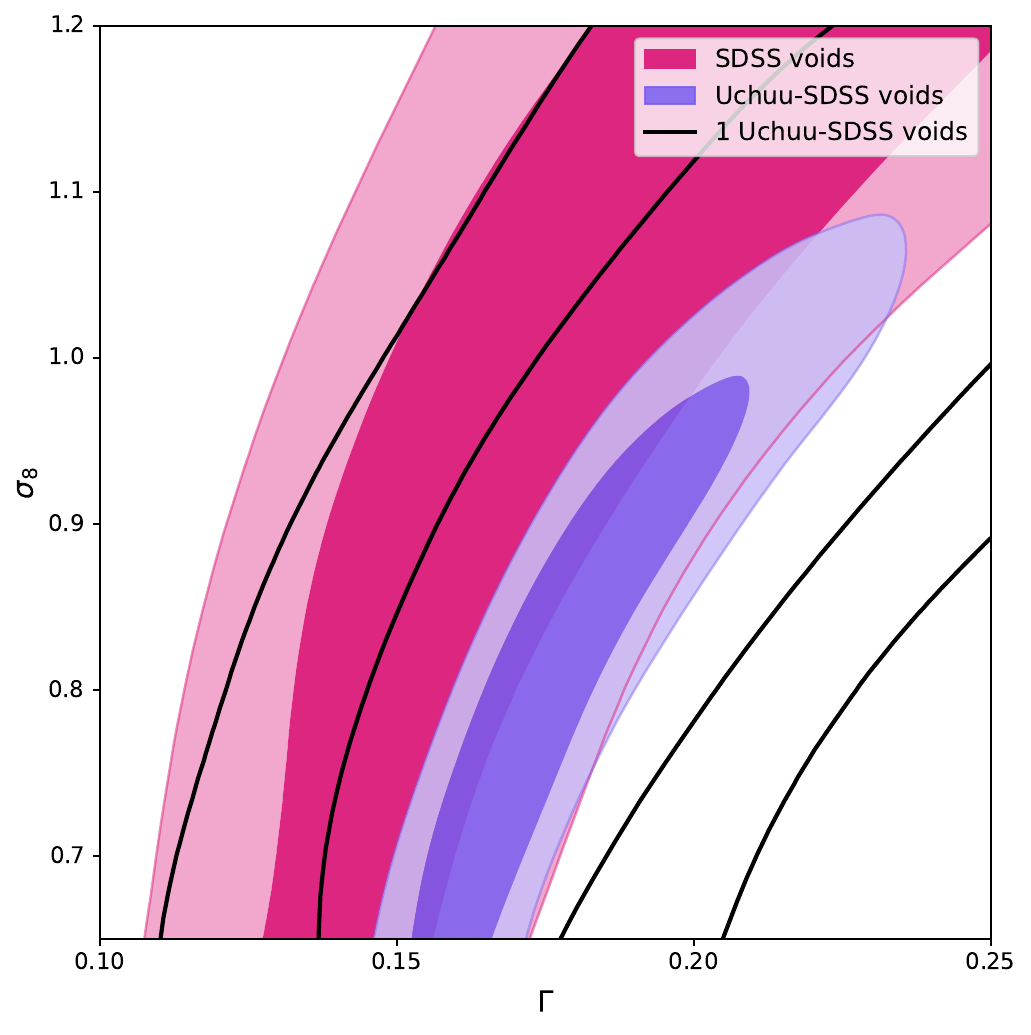}
    \caption{VPF (left panel) and number density of voids larger than $r$ (right panel) predicted by the theoretical framework developed in this work (continuous lines) and measured in Uchuu galaxy box (points) for different galaxy number densities, $n_{g}$, at redshift $z=0.092$.}
    \label{sigma8_gamma_uchuu}
\end{figure}

We can observe the marginalised posterior in the $\sigma_{8}-\Omega_{\rm m}$ plane in Figure \ref{sigma8_gamma_uchuu}. From this figure, it is clear that the marginalised posteriors (and, consequently, the constrained values of each parameter) significantly decrease when a larger sample volume is considered. Additionally, we see that the marginalised posterior for a single random Uchuu-SDSS lightcone void is similar to that obtained for SDSS voids.

In Table \ref{errors_ratio}, we present the constrained values for the voids from a single Uchuu-SDSS lightcone (second column), 5 lightcones (third column), 10 lightcones (fourth column), and 32 lightcones (fifth column) for the main parameters of the theoretical framework: $\sigma_{8}$ and $\Gamma$. The last column shows the power that best fits the relationship $y = x^{1/n}$, where $x$ is the volume ratio between each column (all possible combinations, $V_{2}/V_{1}$ with $V_{1} < V_{2}$) and $y$ is the ratio of uncertainties. We performed two different fits: one for the upper limit uncertainties and another for the lower limit uncertainties. The value of $n$ was averaged from the two fits and then rounded to the nearest integer. However, for $\Gamma$, we obtained $n = -2.5$, so no rounding was applied in this case. The fits were conducted using non-linear least squares.

Therefore, we can observe that the scaling of errors with volume differs for each parameter. The largest scaling is found for $\Gamma$, where the volume scaling is $V^{-1/2.5}$. This means that if we have two samples and one has twice the volume of the other, the errors in the larger sample would be 0.758 times the errors in the smaller sample (i.e. a factor of 1.32 smaller). However, the scaling for $\sigma_{8}$ (and S$_{8}$) is less favourable: $V^{-1/7}$. Using the same example, if one sample has twice the volume of the other, the errors in the larger sample would be 0.906 times the errors in the smaller sample (i.e. a factor of 1.104 smaller).

\begin{table*}
\caption{Difference in the cosmological constraints and their uncertainties using different amounts of Uchuu-SDSS lightcones.}
    \centering
    \begin{tabular}{cccccc}
        \toprule
         &   1 Uchuu-SDSS voids  &5 Uchuu-SDSS voids&10 Uchuu-SDSS voids&32 Uchuu-SDSS voids&$n$\\
         \midrule
         $\sigma_{8}$&   0.997$^{+0.277}_{-0.298}$  &0.886$^{+0.219}_{-0.245}$&0.880$^{+0.199}_{-0.219}$&0.793$^{+0.167}_{-0.187}$&-7\\
 $\Gamma$&  0.2725$^{+0.0924}_{-0.0816}$  &0.2576$^{+0.0605}_{-0.0579}$&0.2203$^{+0.0413}_{-0.0393}$&0.1787$^{+0.0237}_{-0.0217}$&-2.5\\
 \bottomrule
    \end{tabular}
    \tablefoot{Constraints of $\sigma_{8}$ and $\Gamma=\Omega_{\rm c}h$ for voids from 1 Uchuu-SDSS lightcone (second column), 5 Uchuu-SDSS lightcones (third column) 10 Uchuu-SDSS lightcones (fourth column), 32 Uchuu-SDSS lightcones (fifth column) and the index $n$ of the scaling of errors with the volume ($V_{2}/V_{1}^{1/n}$, with $V_{1}<V_{2}$) obtained using non-linear least squares (last column).}
    \label{errors_ratio}
\end{table*}

One way to check the robustness of the values of $n$ is by recalculating the inferences for a different random group of 10 lightcones and comparing the resulting $n$ values. Instead of performing the fit with all cases from Table \ref{errors_ratio}, we consider one lightcone and one group of 10 lightcones, then another lightcone and the other group of 10 lightcones, and finally compare the $n$ values obtained in each case. In the first two columns of Table \ref{errors_n}, we show the inferences obtained for each group (A and B) and the corresponding $n$ values ($n_{A}$ and $n_{B}$). The error in $n$, denoted as $\Delta n$, is calculated as the absolute difference between $n_{A}$ and $n_{B}$.

Therefore, we can conclude that $n_{\sigma_{8}}\sim-7\pm2$ and $n_{\Gamma}\sim-2.5\pm0.1$, so the calculation of $n$, specially for $\Gamma$, is quite robust.

\begin{table*}
\caption{Comparison of the cosmological constraints and their uncertainties using different combinations of 10 lightcones.}
    \centering
    \begin{tabular}{cccccc}
        \toprule
         &   group A&group B&$n_{A}$ &$n_{B}$ &$\Delta n$\\
         \midrule
         $\sigma_{8}$&   0.880$^{+0.199}_{-0.219}$&0.866$^{+0.214}_{-0.238}$&-7.22&-9.58&2.36\\
 $\Gamma$&  0.2203$^{+0.0413}_{-0.0393}$&0.2181$^{+0.0378}_{-0.0399}$&-3.01&-2.87&0.14\\
 \bottomrule
    \end{tabular}
    \tablefoot{Constraints of $\sigma_{8}$ and $\Gamma=\Omega_{\rm c}h$ for voids from two different groups of 10 Uchuu-SDSS lightcones (first and second column, groups A and B respectively) and the index $n$ of the scaling of errors with the volume ($V_{2}/V_{1}^{1/n}$, with $V_{1}<V_{2}$) obtained using non-linear least squares for each groups ($n_{A}$ and $n_{B}$ respectively). $\Delta n$ in the last column is the absolute difference between $n_{A}$ and $n_{B}$.} 
    \label{errors_n}
\end{table*}

\section{Theoretical framework with different number densities of galaxies}
\begin{figure*}[th!]
    \centering
    \includegraphics[width=.99\textwidth]{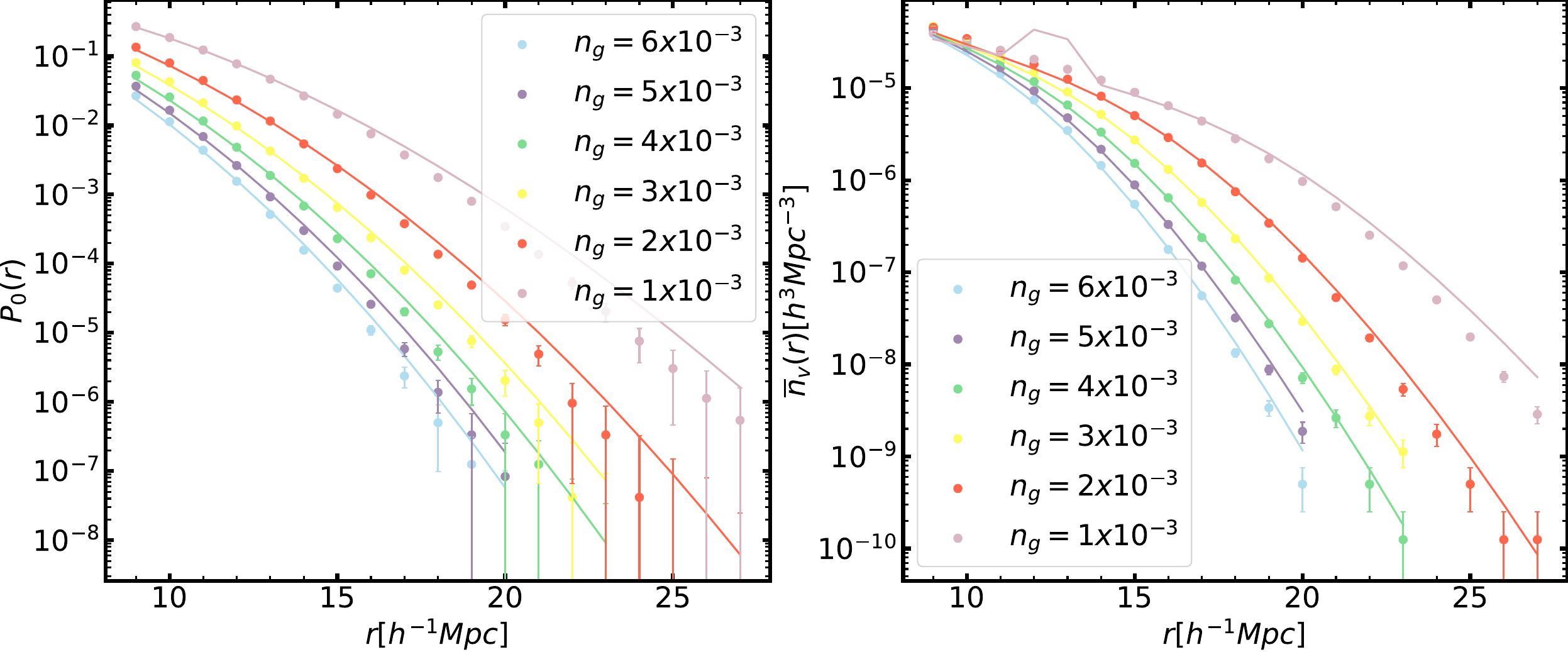}
    \caption{VPF (left panel) and number density of voids larger than $r$ (right panel) in real space predicted by the theoretical framework developed in this work (continuous lines) and measured in Uchuu galaxy box (points) for different galaxy number densities, $n_{g}$ (in units of $h^{3}$Mpc$^{-3}$), at redshift $z=0.092$.}
    \label{theoretical framework_other_densities}
\end{figure*}

In this work, we have constrained the parameters $\sigma_{8}$, $\Omega_{\rm m}$ and H$_{0}$ using galaxy samples with a number density of $3\times10^{-3}h^{3}$Mpc$^{-3}$ at a redshift of $z\sim 0.1$. We have verified that the theoretical framework successfully predicts both the Void Probability Function and the number density of voids larger than $r$ for this number density. However, it remains to be checked whether the theoretical framework continues to successfully predict these two statistics at a different number densities of galaxies. 

To check whether the theoretical framework works correctly for different number densities of galaxies  we need to construct different samples (from Uchuu galaxy simulation box, for example) with different galaxy number densities, calculate the VPF and $n_{v}(r)$ from that sample and compare with the predicted value given by the theoretical framework with the same coefficients that we have calculated in this work. The only parameter that must be changed for each sample is the mass, $m_{g}$. The number densities we use are $n_{g}=\{1\times10^{-3}, 2\times10^{-3}, 3\times10^{-3}, 4\times10^{-3}, 5\times10^{-3}, 6\times10^{-3}\} h^{3}$Mpc$^{-3}$. 

In Figure \ref{theoretical framework_other_densities}, we can observe the VPF (left panel) and number density of voids larger than $r$ (right panel) predicted by theoretical framework with continuous lines and obtained in Uchuu galaxy simulation boxes for these galaxy number densities with dots and the abundance of voids larger than $r$ (right panel). It can be seen that the theoretical framework predicts successfully the VPF for all galaxy number densities without any need to change the values of $\alpha$ or $\mu$, except for galaxy number densities smaller than $n_{g}\leq2\times10^{-3}h^{3}$Mpc$^{-3}$ (see the irregularity obtained for $r$ between 11 and 14 $h^{-1}$Mpc). Therefore, the theoretical framework is only valid for large galaxy number densities ($n_{g}\geq2\times10^{-3}h^{3}$Mpc$^{-3}$), and the coefficients $\alpha$ and $\mu$ depend on this galaxy number density.

\newpage

\section{Content of the void catalogues used in this work}\label{columns}

The columns of the void catalogues of halo and galaxy simulation boxes are the following

\begin{itemize}
    \item X[MPC/H]: x-position of the centre of the void (comoving $h^{-1}$Mpc).
    \item Y[MPC/H]: y-position of the centre of the void (comoving $h^{-1}$Mpc).
    \item Z[MPC/H]: z-position of the centre of the void (comoving $h^{-1}$Mpc).
    \item RADIUS[MPC/H]: radius of the void (comoving $h^{-1}$Mpc).
\end{itemize}

For Uchuu-SDSS lightcones and SDSS, there are four additional columns:

\begin{itemize}
    \item RA[DEG]:  right ascension (degrees).
    \item DEC[DEG]:  declination (degrees).
    \item ZOBS:  observed redshift of the centre of the void (accounting for peculiar velocities. The fiducial cosmology used for SDSS voids is Planck 2015 \citep{planck2015}).
    \item completeness: mean completeness of the void, calculated as the mean completeness of all the points uniformly distributed in its volume.
\end{itemize}

\end{appendix}

\end{document}